\documentclass[fleqn,usenatbib]{mnras}

\usepackage[T1]{fontenc}

\usepackage{graphicx}	
\usepackage{amsmath}	
\usepackage{amssymb}	
\usepackage{comment}
\usepackage{natbib}
\usepackage{xcolor}
\usepackage{pdflscape}
\usepackage{newtxtext,newtxmath}

\def\pp{$P_{P}$~}
\def\tp{$\theta_{P}$~}
\def\bpos{B$_\mathrm{POS}$~}
\def\gaia{\textit{Gaia}~}
\def\planck{\textit{Planck}~}
\def\mua{$\mu_{\alpha\star}$~}
\def\mud{$\mu_{\delta}$~}

\title[Magnetic fields and young stellar objects in L1616]{Magnetic fields and young stellar objects in cometary cloud LDN 1616}

 \author[Piyali Saha et al.]{
 Piyali Saha$^{1}$,\thanks{E-mail: s.piyali16@gmail.com}
 Archana Soam$^{2}$,
 Tapas Baug$^{1}$, 
 Maheswar G.$^{2}$,
 Soumen Mondal$^{1}$, and
 Tuhin Ghosh $^{3}$ 
 \\
 $^{1}$Satyendra Nath Bose National Centre for Basic Sciences (SNBNCBS), Salt Lake, Kolkata-700 106, India\\
 $^{2}$Indian Institute of Astrophysics (IIA), Koramangala, Bangalore 560034, India\\ 
 $^{3}$School of Physical Sciences, National Institute of Science Education and Research, HBNI, Jatni 752050, Odisha, India
 }

\date{Accepted XXX. Received YYY; in original form ZZZ}

\pubyear{2022}

\begin{document}
\label{firstpage}
\pagerange{\pageref{firstpage}--\pageref{lastpage}}
\maketitle

\begin{abstract}
LDN 1615/1616 and CB 28 (hereafter, L1616) together form a cometary globule located at an angular distance of about 8$\degr$ west of the Orion OB1 association, aligned roughly along the east-west direction, and showing a distinct head-tail structure. The presence of massive stars in the Orion belt has been considered to be responsible for the radiation driven implosion mode of star formation in L1616. Based on the latest \textit{Gaia} EDR3 measurements of the previously known young stellar objects (YSOs) associated with L1616, we find the distance to this cloud of 384$\pm$5 pc. We present optical polarimetry towards L1616 that maps the plane-of-sky component of the ambient magnetic field (\bpos\!\!) geometry. Based on the proper motion of the YSOs associated with L1616, we investigate their plane-of-sky motion relative to the exciting star $\epsilon$ Ori. Using the \textit{Gaia} EDR3 measurements of the distances and proper motions of the YSOs, we find two additional sources comoving with the known YSOs. One comoving source is HD33056, a B9 star and the other might be a young pre-main sequence star not reported in previous studies. The mean direction of \bpos is found to follow the cloud structure. This could be the effect of dragging of the magnetic field lines by the impact of the ionizing radiation from $\epsilon$ Ori. Based on the pressure exerted on L1616, and the ages of the associated YSOs, we show that it could possibly be the main source of ionization in L1616, and thus the star formation in it.
\end{abstract}

\begin{keywords}
	Techniques: polarimetric, Stars: distances, pre-main-sequence, Proper motions, ISM: clouds, magnetic fields
\end{keywords}




\section{Introduction}\label{sec:int}

Massive O- and B- type stars exhibit a profound impact on the morphology and evolution of the surrounding molecular cloud structures by their powerful radiation and stellar winds. The energetic photons emitted (at rates of $\sim10^{47}-10^{50}$ s$^{-1}$) from these stars quickly ionize the immediate vicinities, resulting in a shock, which propagates away from the ionizing sources. This eventually results non-uniform cloud structures, such as bright-rimmed clouds (BRCs), cometary globules (CGs), elephant trunk nebula etc., either due to the instabilities \citep[e.g.,][]{1979ApJ...233..280G, 1996ApJ...469..171G, 1999MNRAS.310..789W}, or because of the presence of pre-existing dense clumps in the surrounding area \citep[e.g.,][]{1983A&A...117..183R}. The shocks can also trigger the star formation processes taking place in these cloud structures. The shock-compression could possibly compel the inner dense cores to instability and eventually tend to gravitational collapse, and thus triggering the star formation.

CGs are comet-shaped molecular clouds, consisting of a compact, dusty edge similar to a narrow, bright-rimmed head along with a faint elongated tail-like structure extending from the head and directing away from the source of photoionization. These structures were first identified by \cite{1976MNRAS.175P..19H}, based on Science and Engineering Research Council (SERC) IIIaJ Sky Survey plates. CGs are considered to be a part of Bok globules of having a size range of $\sim0.1-1.0$ pc, often found to be located at the periphery of the H{\sc ii} regions \citep{1976MNRAS.175P..19H, 1983ApL....23..119Z}. They show high number densities, $10^{4} -10^{5}$ cm$^{-3}$ \cite{2007A&A...466..191H, 1994ApJ...433...96V, 1995MNRAS.276.1067B}, with a mass range of $10-100$ M$_{\odot}$ \citep{1994A&A...289..559L, 2007A&A...466..191H}. A number of CGs are found to be sites of ongoing low-mass star formation \citep[e.g.,][]{1983A&A...117..183R, 1998AJ....116.1376S, 2002AJ....123.2597O, 2007MNRAS.379.1237M, 2008ApJ...673..331G, 2013AJ....145...15R}. Several authors discussed the procedures involved in the formation and evolution of CGs in detail \citep[e.g.,][]{1989ApJ...346..735B,1990ApJ...354..529B, 1994A&A...289..559L, 2003csss...12..799K, 2006MNRAS.369..143M,2007IAUS..237..450M}.

\begin{figure*}
	\centering
	\includegraphics[height=8cm,width=16.5cm]{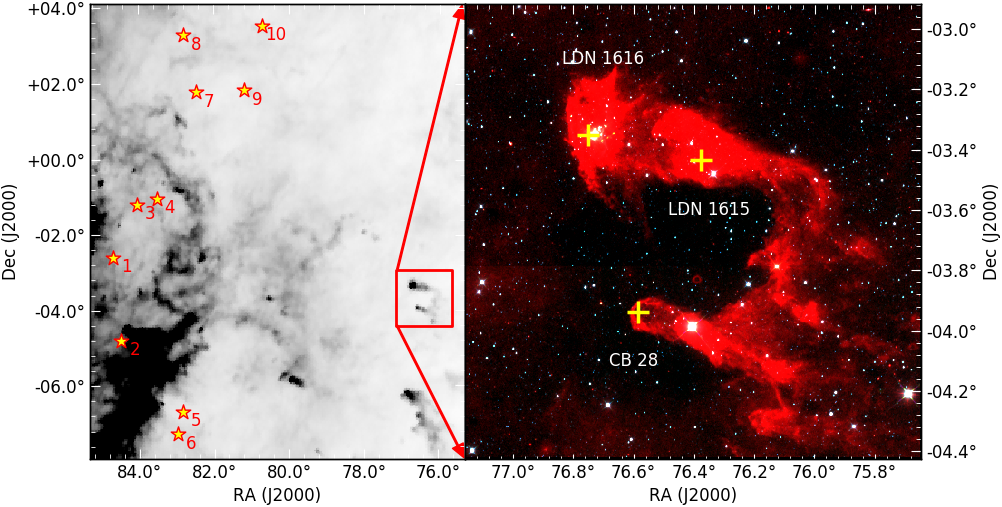}
	\caption{\textbf{(a)} Positions of the bright Orion stars (also listed in Table \ref{tab:ob_stars_table}) near L1616 overplotted on \textit{Planck} $10\degr\times12\degr$ 857 GHz image. Area of analysis of L1616 is indicated by a red colored box of size $1.5\degr\times1.5\degr$. \textbf{(b)} Zoomed in view of the area of analysis is shown, which is \textit{WISE} color-composite image of the L1616 cloud complex using 3.5 (blue), 4.6 (green) and 12 (red) $\mu$m images of size $1.5\degr\times1.5\degr$. The locations of L1616, L1615, and CB28 are marked by yellow colored plus symbols.}
	\label{fig:search_area}
\end{figure*}

The evolution of these CGs and the star formation in them are found to be regulated by the orientation and strength of the ambient magnetic fields \citep{1976ApJ...210..326M, 2000ApJ...540L.103B, 2008A&A...477....9H}. Based on the three-dimensional (3D) magnetohydrodynamical simulations, along with the presence of photoionizing radiation, the impacts of magnetic fields of different orientations and strengths on the formation and evolution of dense globules located at the edges of H{\sc ii} regions were studied  \citep[e.g., ][]{2009MNRAS.398..157H, 2011MNRAS.412.2079M}. They found that to alter the non-magnetized dynamics of the globule significantly, an initially strong magnetic field is necessary, because the photoionizing radiation is strong enough to control the entire dynamics. An initially perpendicular magnetic field of weak and medium strengths, is dragged and made to orient following the cloud structure during the dynamical evolution of the globules. However, in the case of a strong perpendicular field, remains in its initial alignment during the same. Therefore, an investigation of the magnetic field geometry projected on the plane-of-sky (\bpos\!\!) surrounding the molecular clouds could yield major insight into the contribution of the magnetic field in carving the structure and the dynamics of the globules.

The Lynds$^{'}$ Dark Nebula LDN 1615/1616 and CB 28 (hereafter, L1616) resides at an angular distance of about $8\degr$ west of the Orion OB1 associations, centered on the Galactic coordinate $l=203.5384\degr$ , $b=-24.6870\degr$ \citep{1986ApJ...303..375M, 2002A&A...393..251S, 2004A&A...416..677A, 2005ApJ...624..808L, 2007ApJ...657..884L, 2008A&A...491..515C}. This cometary cloud extends about 40\arcmin \citep{2004A&A...416..677A} in the sky and shows evidence of ongoing star formation activity \citep[e.g.,][]{1995PASJ...47..889N,  2002A&A...393..251S, 2004A&A...416..677A}. The head (L1616) points towards the east, which is the general direction of the Orion OB1 associations. L1616 harbours a bright reflection nebula, NGC 1788 (also known as CED 040), which is excited by a star cluster \citep{2002A&A...393..251S}. The brightest visible members of NGC 1788 are two intermediate-mass stars, Kiso A-0974 15 (B3e) and HD 293815 (B9V) \citep{2008ApJ...687.1303G}. The reflection nebula has been considered to be unique compared to the others as its star formation efficiency is $\sim$14\%, which is substantially higher than that typically seen in other clouds associated with reflection nebulae \citep{1995MNRAS.276..923R}. \cite{1986ApJ...303..375M} made a survey of CO (J $=1-0$) line emission towards molecular clouds located at Orion and Monoceros regions. In their survey, L1616 was identified as clump number 13, where a local standard of rest (LSR) radial velocity of 7.7 km s$^{-1}$ (corresponding heliocentric radial velocity was measured as 22.6 km s$^{-1}$) was estimated. The Orion clouds \citep{1986ApJ...303..375M} and the $\lambda$ Ori region \citep{2001AJ....121.2124D} also share similar heliocentric radial velocity ($\sim$25 km s$^{-1}$), which suggests that in spite of the huge angular distance between L1616 and the $\lambda$ Ori region and the Orion molecular clouds, they share a similar radial motion.
	
Based on the RASS ((ROSAT All-Sky Survey)) observations, \cite{1995A&A...297..418S} detected a concentration of X-ray emission in NGC 1788, indicating the presence of star forming regions there. The existence of more pre-main-sequence (PMS) sources was revealed by the Kiso H$\alpha$ survey \citep[][and references therein]{1995PASJ...47..889N} and optical identification of the sources in Orion identified by RASS \citep{2000A&A...353..186A}. Based on millimeter and mid-infrared (MIR) observations, \cite{2002A&A...393..251S} identified five YSOs towards L1616. One among them is presumably a Class 0 YSO, associated with a protostellar jet. \cite{2004A&A...416..677A} made a detailed multiwavelength study towards L1616, by optical spectroscopic and X-ray observations, along with near-IR and optical wide-field imaging observations. They suggested a trend of small-scale sequential star formation scenario towards L1616, with older population towards the north-eastern edge (direction to the Ori OB1 association) of the cloud and younger population towards the head (L1616) region. Therefore, the spatial distribution of the classical- and weak-line T Tauri stars (CTTS and WTTS, respectively), the evidence of multiple star formation events, the location of the nebula near its edge facing the massive ionizing stars, the higher star formation efficiency, and also its comet-like structure altogether support the hypothesis of the star formation in L1615/L1616 being likely triggered by the massive stars in the nearby Orion OB1 associations \citep{1986ApJ...303..375M, 1995MNRAS.276..923R, 2000A&A...353..186A, 2002A&A...393..251S, 2004A&A...416..677A, 2004A&A...418...89K, 2005ApJ...624..808L, 2007ApJ...657..884L}. The luminous stars in Orion OB1 association, which have been considered to be responsible for the cometary shape of the cloud and triggering star formation in L1616 cloud complex \citep{1995MNRAS.276..923R}, are shown in Fig. \ref{fig:search_area} (a) and listed in Table \ref{tab:ob_stars_table}.

In this paper, we attempt to identify the triggering source based on the dynamical study of the YSOs using the recent \gaia early data release 3 (EDR3) measurements. We perform optical polarization observations of this region to understand the magnetic field morphology in the plane-of-the-sky (\bpos\!\!). We also compare our results with the \textit{Planck} polarization measurements towards this cloud. Based on the \textit{Gaia} EDR3 data, we make an attempt to find the possible acceleration of the associated YSOs with respect to the ionizing source on the plane-of-sky. Based on the magnetic field geometry of L1616 and the relative proper motion of the YSOs, we investigate the possible radiation-driven implosion (RDI) towards this cloud. We additionally find several sources sharing similar distance and proper motion as the previously known YSOs. This paper is organized in the following manner: Section \ref{sec:obs} presents the observational and archival data used in this work. In Section \ref{sec:res}, we describe the results and the relevant discussion. Finally, we conclude this paper by summarizing the results in Section \ref{sec:con}.

\section{Observations, Archival Data \& Data Reduction}\label{sec:obs}

\subsection{Observational Data} \label{sec:aimpol}

In order to map the \bpos towards L1616, the optical polarimetric observations have been carried out using Aries IMaging POLarimeter \citep[AIMPOL;][]{2004BASI...32..159R} mounted as a back-end instrument at the f/13 Cassegrain focus of the 104 cm Sampurnanand Telescope, located at Aryabhatta Research Institute of observational SciencES (ARIES), Nainital, India. Table \ref{tab:log_table} lists the log of observations. Description of the instrument and the methods of the polarization measurements are provided in detail in \cite{2004BASI...32..159R}. The polarimetric data reduction steps and the procedures adopted to estimate the polarization fraction ($P$) and position angle ($\theta$) of the observed stars are discussed in several previous studies \citep[e.g., ][]{2011MNRAS.411.1418E, 2013MNRAS.432.1502S, 2021A&A...655A..76S}. The observations have been carried out using the $R_{c}$ photometric band ($\lambda_{R_{eff}}=0.670~ \mu$m) using the central area of $370\times370$ pixel$^{2}$ of the TK $1024\times1024$ pixel$^{2}$ CCD camera.

\begin{table}
	\begin{center}
	\caption{Luminous stars close to L1616.}
	\label{tab:ob_stars_table}
	\renewcommand{\arraystretch}{1.5}
	\begin{tabular}{lcccr} 
		\hline
		Star  & Star & Distance & Distance from & Spectral \\
		ID &  Name & (pc) &  L1616$^{\ddagger}$ (pc) & Type$^{*}$ \\
		\hline
		1	& HD 37468$^{\dagger}$  & 352$_{-113}^{113}$ & 60$_{-53}^{53}$ &O9.5V\\
		2	& HD 37356  & 448$_{-10}^{12}$ & 86$_{-9}^{10}$ &B2IV-V\\
		3	& HD 37128$^{\dagger}$  & 412$_{-154}^{154}$ & 60$_{-81}^{81}$ &B0Ia\\
		4	& HD 36779  & 367$_{-11}^{13}$ & 50$_{-4}^{4}$ &B2.5V\\
		5	& HD 36430  & 393$_{-10}^{11}$ & 48$_{-3}^{3}$ &B2V\\
		6	& HD 36512  & 407$_{-24}^{25}$ & 56$_{-11}^{12}$ &O9.7V\\
		7	& HD 36166  & 342$_{-11}^{15}$ & 64$_{-7}^{10}$ &B2V\\
		8	& HD 36351  & 323$_{-11}^{12}$ & 82$_{-8}^{9}$ &B1.5V\\
		9	& HD 35439  & 342$_{-12}^{13}$ & 60$_{-9}^{9}$ &B1Vpe\\
		10	& HD 35149$^{+}$  & 575$_{-78}^{119}$ & 202$_{-75}^{115}$ &B2\\
		\hline
	\end{tabular}\\
	\end{center}
	$^{\dagger}$ The \textit{Gaia} measurements of HD 37468 and HD 37128 are not available. Hence, distance is derived from parallax information provided by \textit{Hipparcos} archive.\\
	$^{\ddagger}$ True distances between the luminous stars and L1616.\\
	$^{*}$ Spectral types were obtained from Simbad database. \\
	$^{+}$ RUWE of HD 35149 is $> 1.4$, making its astrometric measurements unreliable. 	
\end{table}

Unpolarized standard stars have also been observed during each run to check for the instrumental polarization. The typical instrumental polarization for this set up is found to be $\sim0.1$\% \citep{2004BASI...32..159R, 2008MNRAS.388..105M, 2011MNRAS.411.1418E, 2014Ap&SS.350..251S, 2016A&A...588A..45N, 2018MNRAS.476.4782S}. The final polarimetric results are corrected for instrumental polarization. To obtain the reference direction of the polarizer, a few polarized standard stars (BD+59${\degr}$389, HD 19820, HD 236633, and HD 25443) have been selected from \cite{1992AJ....104.1563S} and observed on each observing run. We estimate the zero point offset of the polarizer by using the difference between the computed $\theta$ of individual polarized standard star and its value provided by \cite{1992AJ....104.1563S} on every observing run.

\begin{table}
	\centering
	\caption{Log of observations towards L1616 using R$_{c}$ filter.}
	\label{tab:log_table}
	\begin{tabular}{ll}
		\hline
		year & month (date)\\
		\hline
		2012 & January (31), February (26), March (29)\\
		2013 & February (9), December (1, 2, 3, 28)\\
		2014 & January (25, 26), February (5)\\
        2016 &	February (5, 6, 8, 9), October (25, 27)\\        		
        2017 &	March (21), October (14, 20, 21, 26, 27)\\
        \hline
	\end{tabular}\\
\end{table}

\subsection{Archival Data} \label{sec:gaia}

\subsubsection{\textit{Gaia} Early Data Release 3}

The European Space Agency mission \textit{Gaia} Early Data Release 3 (EDR3) \citep{2021A&A...649A...1G} provides astrometric measurements of 1.8 billion objects with unprecedented precision, brighter than $G$-band magnitude $\sim$21. The \textit{Gaia} EDR3 catalog contains results based on data acquired during the first 34 months of the mission, resulting in improvements in the photometric and astrometric measurements because of the enhanced number of raw observations. For brighter ($G<15$) sources the typical uncertainties in parallax are $0.02-0.03$ mas, while the same for fainter sources ($G=21$) are 1.3 mas. Also, for brighter ($G<15$) sources the typical uncertainties in proper motion are $0.02-0.03$ mas yr$^{-1}$, while the same for fainter sources ($G=21$) are 1.4 mas yr$^{-1}$ \citep{2021A&A...649A...1G}. We obtain distance (\textit{d}) and proper motions in right ascension ($\mu_{\alpha\star}$) and declination ($\mu_{\delta}$) from \cite{2021AJ....161..147B} and the \textit{Gaia} EDR3 database \citep{2021A&A...649A...1G} by making a search around each source within a search radius of 1$\arcsec$. Only sources with m/$\Delta$m $\geqslant3$ are considered in this study, where m represents the \textit{d}, $\mu_{\alpha\star}$ and $\mu_{\delta}$ values and the $\Delta$m represents their respective mesurement uncertainties. We exclude sources for which renormalized unit weight error (RUWE)$>$1.4 \citep{LL:LL-124}.

\begin{figure}
	\includegraphics[height=7.7cm,width=\columnwidth]{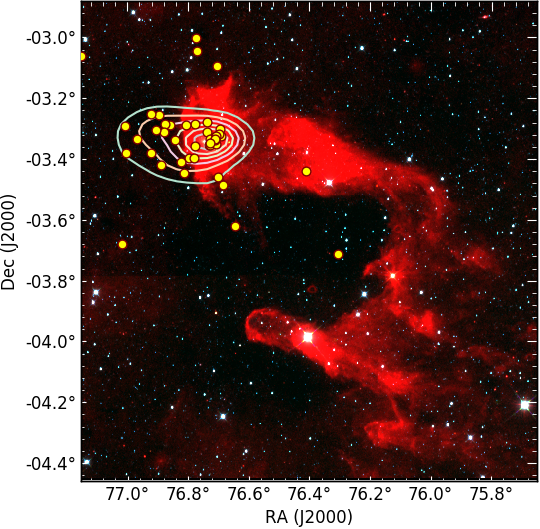}
	\caption{The known YSOs distributed towards L1616 obtained from literature are shown using filled yellow circles on the \textit{WISE} color-composite image of L1616 using 3.5 (blue), 4.6 (green) and 12 (red) $\mu$m images of size $1.5\degr\times1.5\degr$. The contours shown in thick lines represent the surface density distribution of these YSOs in this region.}
	\label{fig:l1616_yso_level}
\end{figure}

\subsubsection{\textit{Planck} polarization measurements in sub-millimetre}

\textit{Planck} built the first all-sky map of the polarized dust emission at sub-millimeter wavelengths \citep{2016A&A...594A...1P}. In order to obtain the polarization fraction and position angle in \textit{Planck} (\pp and \tp\!, respectively), we have used Stokes $I$, $Q$, and $U$ maps at 353 GHz only as this is the highest frequency polarization-sensitive channel and has the best signal to noise ratio for polarized thermal dust emission \citep{2015A&A...576A.104P}. In this study, the whole sky map at 353 GHz (bandpass leakage corrected) is used, which is provided by \textit{Planck} Legacy Archive\footnote{\url{http://www.cosmos.esa.int/web/planck/pla/}}. Also, the dust emission is $\sim2$ orders of magnitude higher than the polarized cosmic microwave background (CMB) at the same frequency \citep{2016A&A...594A...1P}. Therefore, we have not corrected for the CMB polarization in the analysis. At 353 GHz, the angular resolution of \textit{Planck} data is $\sim4.8\arcmin$. The Stokes parameter maps are built using the gnomonic projection of the HEALPix\footnote{\url{http://healpix.sourceforge.net}} all-sky maps \citep{2005ApJ...622..759G}. We have used {\it healpy} module in python \citep{Zonca2019} to extract and analyze the $I$, $Q$, and $U$ maps to estimate the \pp and \tp \footnote{\textit{healpy} is a Python package to handle pixelated data on the sphere.}. We have acquired the Stokes $I$, $Q$, and $U$ maps from the smoothed \textit{Planck} map at the \textit{Planck} 353 GHz frequency. We correct the maps for the cosmic infrared background (CIB) monopole by subtracting 0.13 MJy sr$^{-1}$ (or 452 $\mu$K$_\mathrm{CMB}$, using the unit conversion 287.5 MJy sr$^{-1}$K$_\mathrm{CMB}$) from the maps \citep{2020A&A...641A..12P}. To account for the systematic effect related to the warm ionized medium (WIM)-associated dust, along with the Galactic H{\sc i} offset, 63 $\mu$K$_\mathrm{CMB}$ is added to the maps \citep{2020A&A...641A..12P}. The Stokes $I$, $Q$, and $U$ maps are presented in the IAU (International Astronomical Union) convention, where the polarization angle $\psi$ is 0$\degr$ towards the Galactic north, and increases towards the Galactic east \citep{refId0}. The angle of \bpos ($\chi$) can be estimated by adding 90$\degr$ to the polarization angle ($\chi=$\tp$+90\degr$).

\subsubsection{H{\sc i} 4$\pi$ (HI4PI) Survey}

The H{\sc i} 4$\pi$ (HI4PI) Survey \citep{2016A&A...594A.116H} is a 21-cm all-sky survey of neutral atomic hydrogen. The Effelsberg-Bonn H{\sc i} Survey (EBHIS) and the Galactic All-Sky Survey (GASS) were combined to develop this survey. EBHIS observations were performed with the 100-m radio telescope located at Effelsberg/Germany, while GASS observations were conducted with the Parkes 64-m telescope. The angular resolution of HI4PI Survey is 16.2$^{\prime}$ and sensitivity 43 mK. We present the HI4PI image of L1616 to present the H{\sc i} line emission from this cloud. The HI4PI data have been retrieved from the CDS (Centre de Donn$\mathrm{\acute{e}}$es astronomiques de Strasbourg) via \url{ http://cdsarc.u-strasbg.fr/viz-bin/qcat?J/A+A/594/A116}.

\begin{figure}
	\includegraphics[height=7cm,width=\columnwidth]{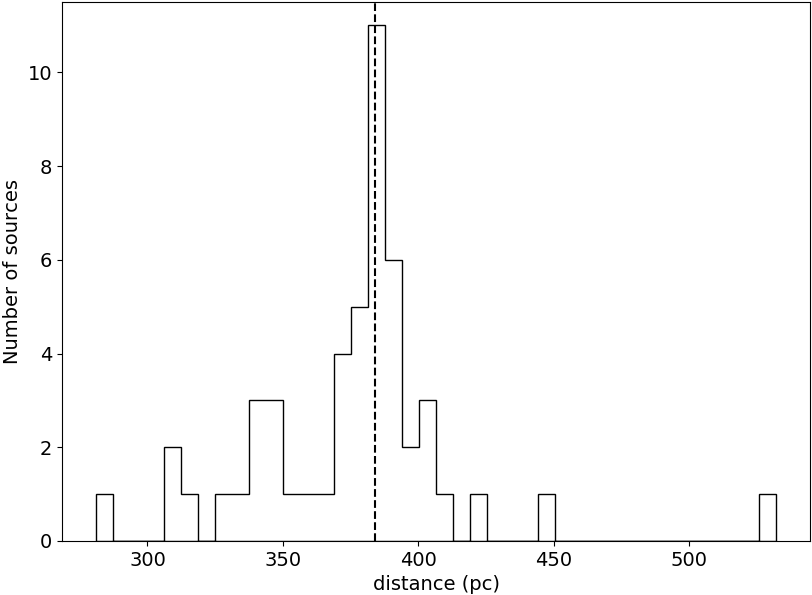}
	\caption{Histogram of distances of the known YSOs acquired from \textit{Gaia} EDR3 constructed for a bin size of $\sim7$ pc. The vertical dashed line represents the median value of the $d$.} 
	\label{fig:l1616_hist_d}
\end{figure}

\begin{figure}
	\includegraphics[height=6.5cm,width=\columnwidth]{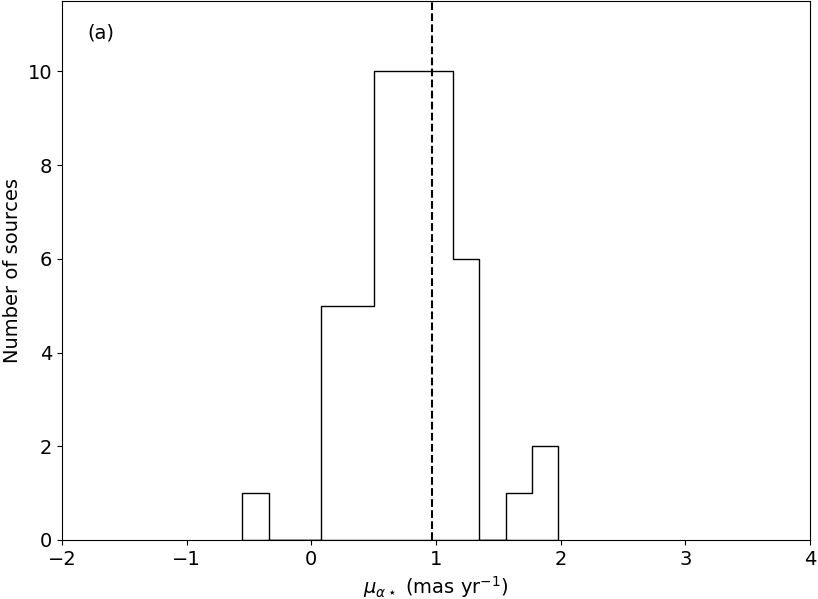}
	\includegraphics[height=6.5cm,width=\columnwidth]{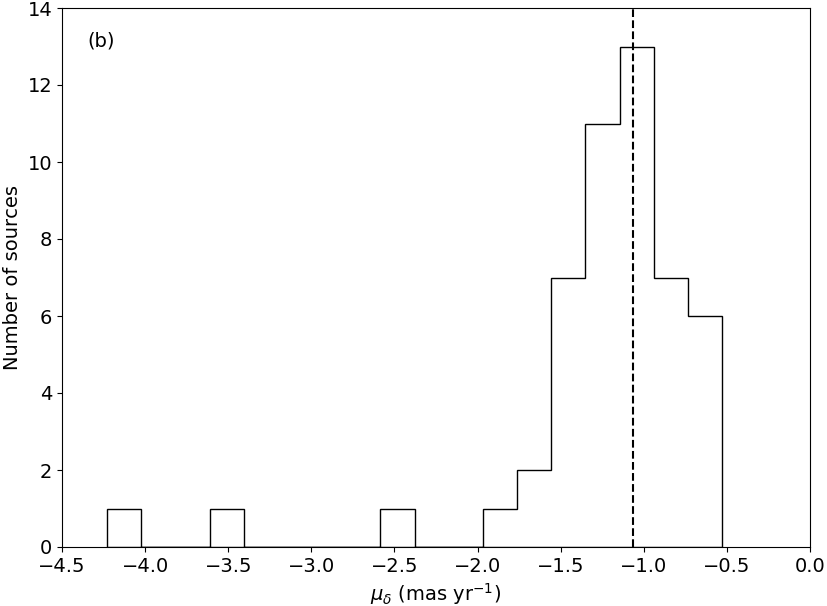}
	\caption{Distribution of proper motions of the known YSOs. \textbf{(a)} Histogram of \mua of the known YSOs acquired from \textit{Gaia} EDR3. The dashed line presents the median value of the \mua\!. Bin size of the histogram $\sim0.2$ mas yr$^{-1}$. \textbf{(b)} Histogram of \mud of the known YSOs acquired from \textit{Gaia} EDR3. The vertical dashed line indicates the median value of the \mud. Bin size of the histogram $\sim0.2$ mas yr$^{-1}$.} 
	\label{fig:l1616_hist_pmdec}
\end{figure}

\begin{figure}
	\includegraphics[height=9.7cm,width=\columnwidth]{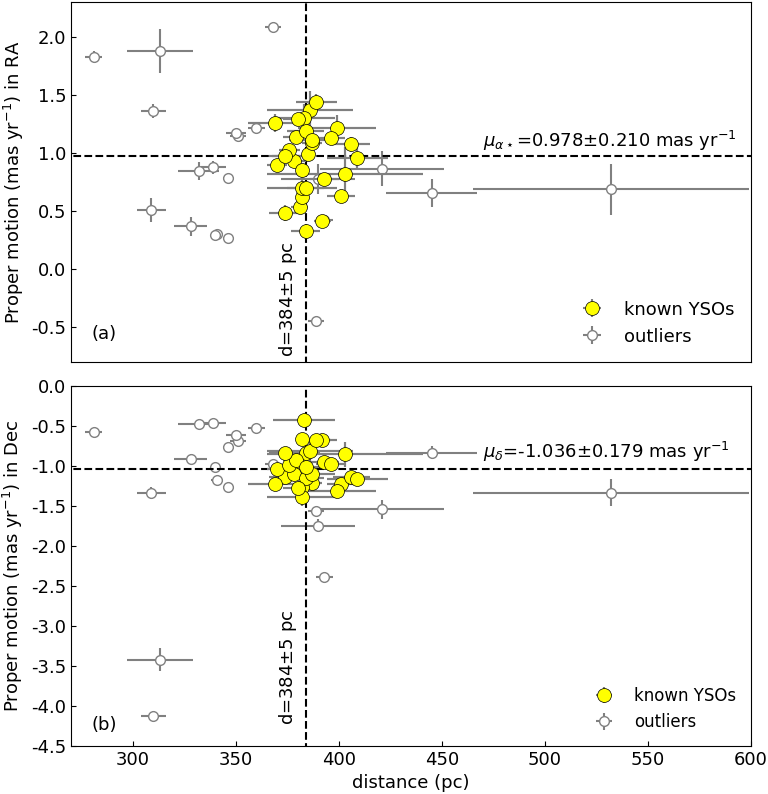}
	\caption{Distribution of proper motions of the known YSOs as a function of their distances acquired from \textit{Gaia} EDR3. The $d-\mu_{\alpha\star}$ and $d-\mu_{\delta}$ values of the YSOs, lying within 3$\times$ MAD boundaries, are shown using the yellow filled circles in (a) and (b), respectively. The outliers are shown using open circles. The dashed lines show the median values of $d$, $\mu_{\alpha\star}$, and $\mu_{\delta}$.}
	\label{fig:l1616_d}
\end{figure}

\section{Results and discussion:}\label{sec:res}

\subsection{Estimation of the distance to L1616}

\cite{1968AJ.....73..233R} estimated a distance to HD 293815, associated with the reflection nebula to be $\sim420$ pc. \cite{1985Sci...230.1150H} provided the distance to L1616 as 420 pc. \cite{1995MNRAS.276..923R} also adopted the distance to L1616 as 420 pc in the molecular line analysis. \cite{2004A&A...416..677A} considered the distance to Orion as the same to L1616, which is 450 pc. Later, \cite{2008hsf1.book..732W} estimated a distance to L1616 as 420$\pm$30 pc, while reviewing the distance to the $\sigma$ Orionis cluster. In order to study the internal motions in OB associations using \textit{Gaia} data release 2 (DR2), \cite{2020MNRAS.493.2339M} adopted a distance to Ori OB1 as 400 pc, where L1616 is located. Recently, based on the \textit{Gaia} EDR3 data, \cite{2021ApJ...917...21S} estimated a distance to Ori OB1 as $\sim$355 pc.

\begin{figure}
	\includegraphics[width=\columnwidth]{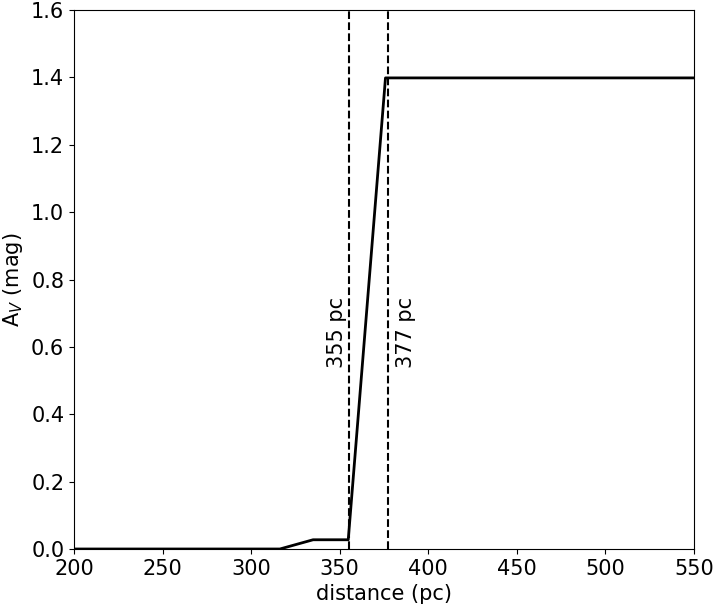}
	\caption{Visual extinction $A_\mathrm{V}$ towards L1616 as a function of the distance. The vertical dashed lines are 355 and 377 pc, where the $A_\mathrm{V}$ changes sharply with the distance.}
	\label{fig:l1616_av_d}
\end{figure}

In this work, we determine the distance to L1616 using the YSOs that are projected towards the cloud. A similar approach was made in several studies \citep[e.g.,][]{2018ApJ...867..151D, 2018ApJ...865...73O, 2020MNRAS.494.5851S, 2022MNRAS.510.2644S}. We have collected the YSOs located towards L1616 from the literature \citep{2002A&A...393..251S, 2004A&A...416..677A, 2008ApJ...687.1303G, 2018A&A...620A.172Z} within an area of 1.5$\degr$$\times$1.5$\degr$ centered on ($\alpha=76.401368\degr$, $\delta=-3.6698349\degr$). The choice of such region iss based on our polarimetric observation towards this area. A total of 65 YSOs have been identified towards the defined region, which is shown in Fig. \ref{fig:l1616_yso_level}. The thick contours represent the source distribution around L1616. As evident from Fig. \ref{fig:l1616_yso_level}, the distribution of the YSOs is not symmetric around the central reflection nebula; rather, they are distributed in an elongated manner towards the massive ionizing sources in the east-west direction. Of the 65 YSOs, we obtain distance and proper motion values in the right ascension ($\mu_{\alpha\star}=\mu_{\alpha}$cos$\delta$) and declination ($\mu_{\delta}$) for 50 sources from \cite{2021AJ....161..147B} and \gaia EDR3 \citep{2021A&A...649A...1G} catalogues, respectively, which have m/$\Delta$m$\geq$3 and RUWE $\leq$1.4. The histogram of $d$ of the YSOs is shown in Fig. \ref{fig:l1616_hist_d}. The median value of $d$ is estimated as 382 pc, which is marked with a vertical dashed line. Similarly, Fig. \ref{fig:l1616_hist_pmdec} (a) and (b) present the histograms of \mua and \mud, respectively. The median values of \mua and \mud are estimated as $0.887$ and $-1.027$ mas yr$^{-1}$, respectively, which are marked with vertical dashed lines. The proper motion measurements (\mua and \mud) of these YSOs as a function of their distances are shown in Fig. \ref{fig:l1616_d} (a) and (b), respectively. A clustering of sources is clearly depicted in Fig. \ref{fig:l1616_d}. Now, in order to estimate the statistical dispersion in these parameters, we used the median absolute deviation (MAD) because MAD is more resilient to outliers in a data set compared to the standard deviation. The MAD in \mua\!\!, \mud\!\!, and $d$ are computed to be $0.261$ mas yr$^{-1}$, $0.209$ mas yr$^{-1}$, and $13$ pc, respectively. In Fig. \ref{fig:l1616_d} (a) and (b), we can see that a majority of the YSOs (29) are falling within the constraints of 3 $\times$ MAD in proper motions and distances, with respect to the median values of the \mua\!\!, \mud\!\!, and $d$, which are shown in filled yellow circles. The results of these sources are listed in Table \ref{tab:YSO_gaia}. The sources that are located outside of the 3 $\times$ MAD boundaries are considered to be outliers. These outliers are the sources that might not be associated kinematically with the cluster in L1616. Excluding these outliers, the median of \mua\!\!, \mud\!\!, and $d$ of the YSOs are estimated as $0.978$ mas yr$^{-1}$, $-1.036$ mas yr$^{-1}$, and $384$ pc, respectively, and the MAD in same are found to be $0.210$ mas yr$^{-1}$, $0.179$ mas yr$^{-1}$, and $5$ pc, respectively. Thus, based on the YSOs associated with L1616, we estimate the distance to the cloud as $384\pm5$ pc. 

We obtain the visual extinction $A_\mathrm{V}$ versus the distance plot of L1616 from the 3D maps of \cite{2019ApJ...887...93G}. Here we have used the Python based `dustmap' package \citep{2018JOSS....3..695M} and the best-fitted model in order to download the reddening values. R$_\mathrm{V} = 3.1$ is used to convert those values $A_\mathrm{V}$. Fig. \ref{fig:l1616_av_d} presents the distribution of $A_\mathrm{V}$ with respect to the distance along the direction of L1616. In this figure, it is clearly visible that there is a change in $A_\mathrm{V}$ from $\sim355-377$ pc. Beyond $\sim377$ pc, the $A_\mathrm{V}$ increases to a higher value of $\sim1.4$. Our estimation of the distance to L1616 as $384\pm5$ pc is consistent with this.

The distance to L1616 as 384$\pm$5 pc indicates that L1616 is located at about 165 pc below the Galactic mid-plane, which is higher than the scale height estimated for the Gould Belt \citep[$\sim$ 56 pc,][]{2020AstBu..75..267B}. Thus, L1616 is an example of star forming cloud consisting of a sparse group of young stars with higher star formation efficiency \citep[$\sim14$\%;][]{1995MNRAS.276..923R} at a relatively distant Galatic latitude.

\begin{figure*}
	\centering
	\includegraphics[height=7.8cm, width=8.5cm]{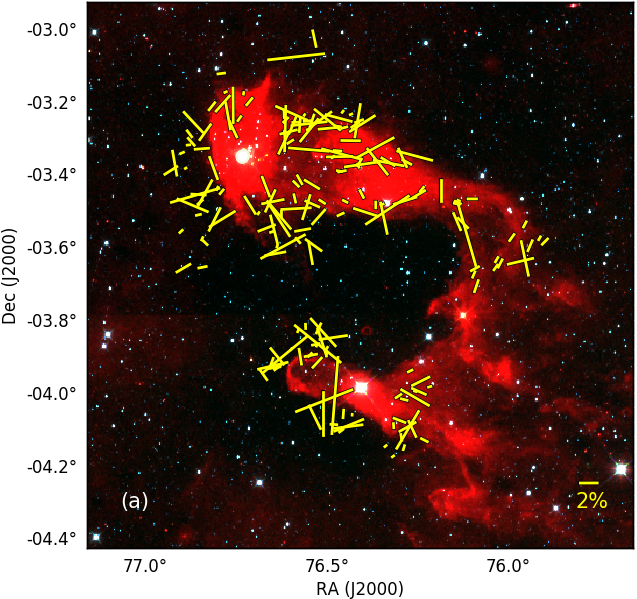}
	\includegraphics[width=8.5cm, height=7.8cm]{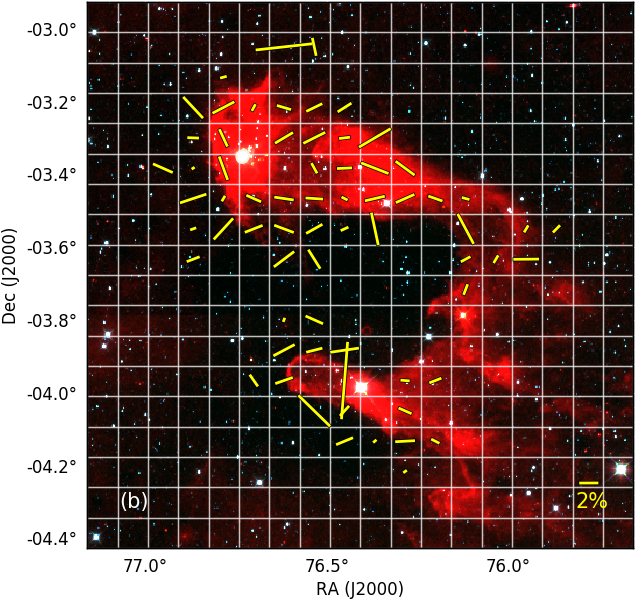}
	\caption{\textbf{(a)} The foreground corrected optical polarization vectors (yellow lines) overplotted on the color-composite image using \textit{WISE} 3.5 (blue), 4.6 (green), and 12 (red) $\mu$m images of $1.5\degr\times1.5\degr$ size. The reference vector with 2\% polarization has been shown. \textbf{(b)} The mean polarizations and position angles of \textbf{(a)} (yellow lines) overplotted on the image same as \textbf{(a)} of L1616. The grids over which the mean values were taken are shown using white colored boxes. The reference vector with 2\% polarization is also shown.}
	\label{fig:pol_rslts_figure}
\end{figure*}

\subsection{Foreground polarization correction}

While proceeding through an ensemble of aspherical dust grains, the unpolarized starlight converts into partially plane polarized light as a consequence of selective extinction. Because of the asphericity of the dust grain, the absorption of stellar radiation happens maximum along the dust grain's long axis, therefore, the transmission would be maximum along the short axis. As the dust grains are prone to align their minor axes parallel to the \bpos because of angular momentum, the transmitted light would follow the direction of the \bpos\!\!. Therefore, the \bpos on the boundaries of dense molecular clouds can be traced using linear polarization measurements of the background stars \citep[e.g.,][]{1990ApJ...359..363G, 2008A&A...486L..13A, 2010ApJ...723..146F}. We use the threshold of $P/\Delta{P}\geqslant2$ in the analysis of $R$-band polarimetric data, where $\Delta{P}$ is the uncertainty in the measurement of $P$. This threshold extracts 195 stars, which are located background of the L1616 cloud complex. The distances of the observed stars are obtained from \cite{2021AJ....161..147B}. We relax our initial selection criteria of $d/\Delta d\geq3$ and RUWE$\leq$1.4 on these 195 sources as here our primary investigation is focussed only on the polarimetric measurements. Therefore, we use polarimetric results of these 195 sources to trace the \bpos geometry around L1616.

\begin{figure}
	\includegraphics[height=7.5cm, width=\columnwidth]{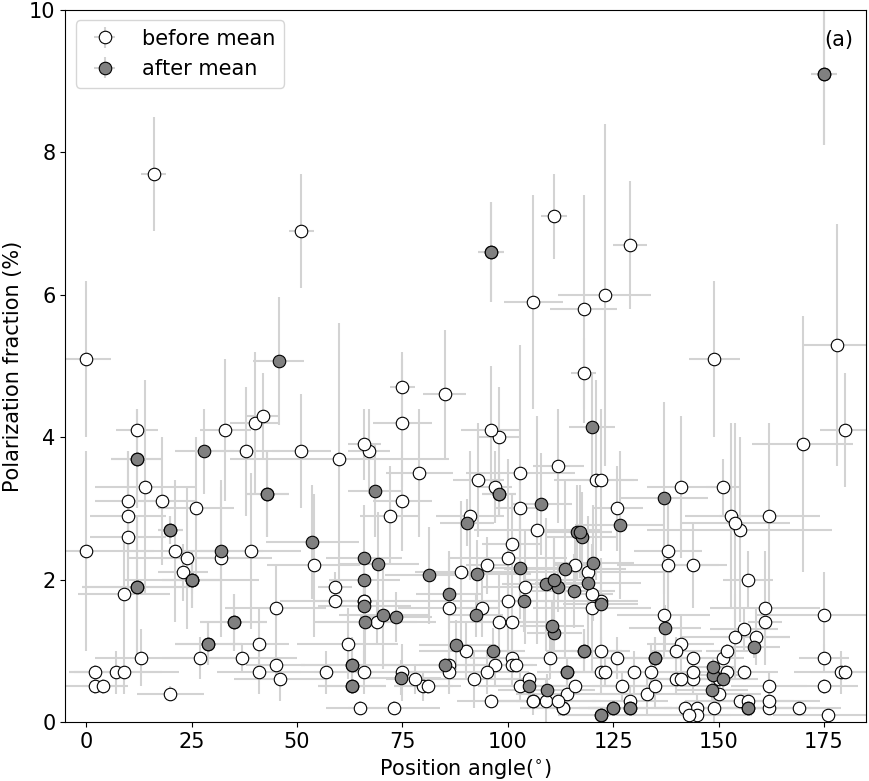}
	\includegraphics[height=7.5cm, width=\columnwidth]{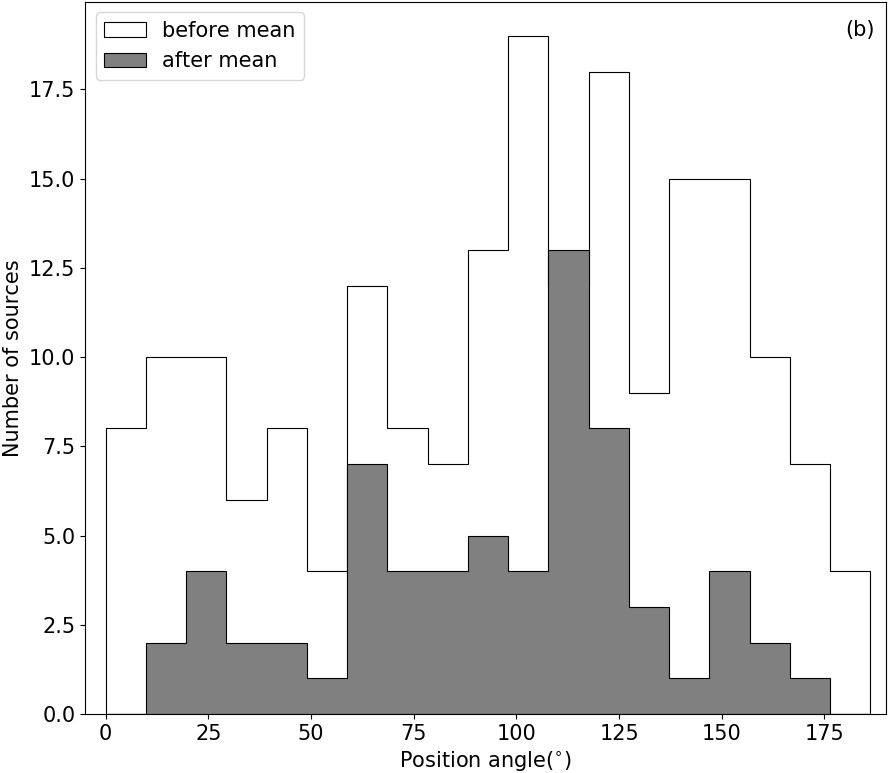}
	\caption{\textbf{(a)} Distribution of $P$ with $\theta$ towards L1616 before and after taking mean are shown using white and gray filled circles, respectively. \textbf{(b)} Histograms of $\theta$ of the stars before and after taking mean are shown using white and grey colors, respectively, with bin size $\sim10\degr$.} 
	\label{fig:p_theta_figure}
\end{figure}

\begin{table}
	\begin{center}	
	\caption{Polarization measurements of the six foreground stars observed in R$_{c}$ band.}
	\label{tab:pol_frg_table}
	\renewcommand{\arraystretch}{1.5}
	\begin{tabular}{|c|c|c|c|c|}
		\hline
		Star & Star & P$\pm\Delta$P & $\theta\pm\Delta\theta$ &  Distance$^{\dagger}$\\
		ID & Name & (\%) & ($^{\circ}$) & (pc)\\
		\hline
		1       & HD 33608  & $0.10\pm0.02$ & $166\pm15$ &  37$_{-0.1}^{0.1}$\\
		2       & HD 33208  & $0.11\pm0.06$ & $112\pm09$ &  89$_{-0.2}^{0.2}$\\
		3       & HD 33345  & $0.10\pm0.05$ & $168\pm09$ &  144$_{-0.5}^{0.5}$\\
		4       & HD 32884  & $0.11\pm0.07$ & $102\pm10$ &  174$_{-0.8}^{0.7}$\\
		5       & HD 32721  & $0.10\pm0.06$ & $108\pm10$ &  178$_{-1.0}^{1.0}$\\
		6       & HD 33023  & $0.24\pm0.09$ & $154\pm08$ &  174$_{-0.8}^{0.9}$\\
		\hline
	\end{tabular}\\
	\end{center}
	$^{\dagger}$ Distances have been adopted from \citet{2021AJ....161..147B}. Uncertainties in distance are shown upto one decimal place.\\
\end{table}

The polarization of the stars located towards a cloud could be the combination of one component by the dust located foreground to the cloud and another component by the dust located in the cloud. To obtain the true polarization of the background stars, which infers the true orientation of the \bpos\!\!, one needs to subtract the foreground contribution from the observed polarization of the stars. Therefore, we have selected six foreground stars, which are distributed within a circular region of $1\degr$ radius about L1616 and observed them in polarimetric mode using AIMPOL. Table \ref{tab:pol_frg_table} presents the polarimetric results of these six stars in the ascending order of their distances. In addition to these observed sources, 19 stars with their polarization results are found to be available in \cite{2000AJ....119..923H} within a circle of radius $5\degr$. The distances of these stars are obtained from \cite{2021AJ....161..147B}. Of the 19 \textit{Heiles} stars, the stars located up to 352 pc are used additionally with the observed foreground stars for the removal of foreground polarization.

The mean values of $P_{fg}$ and $\theta_{fg}$ of all the foreground stars values are found to be 0.15\% and $107\degr$, respectively. The corresponding Stokes parameters $Q_{fg}(=P$cos2$\theta$) and $U_{fg}(=P$sin2$\theta$), are estimated to be $-0.126$ and $-0.083$, respectively. First, the Stokes parameters of the observed background stars ($Q_{\ast}$ and $U_{\ast}$) are computed. Then, the foreground-corrected Stokes parameters of the observed sources ($Q_{c}$ and $U_{c}$) are estimated using the equations $Q_{c}=Q_{\ast}-Q_{fg}$ and $U_{c}=U_{\ast}-U_{fg}$, respectively. The corresponding foreground-corrected polarization fraction ($P_{c}$) and the position angle ($\theta_{c}$) of the observed stars are estimated using the equations $P_{c} = \sqrt{Q_{c}^{2} + U_{c}^{2}}$ and $\theta=$ 0.5 $\tan^{-1}(U_{c}/Q_{c})$, respectively. Fig. \ref{fig:pol_rslts_figure} (a) shows the polarization vectors (yellow lines) overplotted on \textit{WISE} (Wide-field Infrared Survey Explorer) color-composite image of L1616. The lengths and the orientations of the vectors (starting from the north and increasing eastward) represent the estimated $P$ and $\theta$ values, respectively. The low value of foreground polarization along the direction of L1616 at higher Galactic latitude ($b=-24.6\degr$) is not unexpected. This is evident for relatively low column density along the line-of-sight towards the clouds located at higher latitude. Since $Q$ and $U$ can have both positive or negative values, the polarization $P$ computed using these parameters is always positive. Therefore, $P$ has a positive bias, especially for the sources, which have low signal-to-noise ratio. To remove this positive bias, we estimate the debiased $P$ using the relation $P=\sqrt{P^{2}-\Delta{P}^{2}}$ \citep{1974ApJ...194..249W, 2006PASP..118.1340V}.

\subsection{Orientation of Magnetic Field towards L1616 inferred from $R$-band polarimetric data}

\begin{figure}
	\includegraphics[height=8cm,width=\columnwidth]{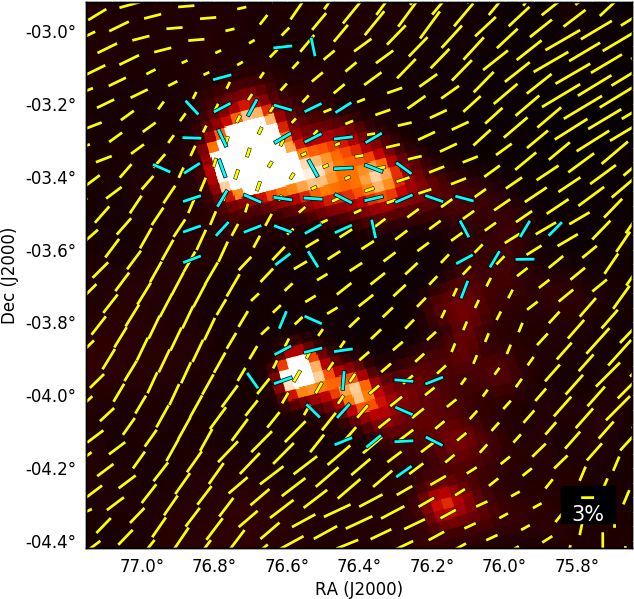}
	\caption{Magnetic field geometry towards L1616 obtained from \textit{Planck} observations are shown using yellow lines, along with observed normalized polarimetric results (after taking mean, using cyan lines) with 3\% polarizations overplotted on $1.5\degr\times1.5\degr$ \textit{Planck} 857 GHz image.}
	\label{fig:planck_figure}
\end{figure}

In Fig. \ref{fig:pol_rslts_figure} (a), the magnetic field lines traced by our $R$-band polarization measurements seem to be random. However, as we move westward, the field lines follow the curved structure of the tail. Table \ref{tab:pol_rslt_table} in appendix presents our foreground-corrected optical polarimetric results of 195 stars located at the background of L1616. The right ascensions and declinations of the sources are listed in columns (2) and (3), respectively. The $P$ and $\theta$ of the observed sources are provided in columns (4) and (5), respectively. The range of estimated $P$ and $\theta$ are $0.1-9.1$\% and $0\degr-180\degr$, respectively. The dispersion values of the same are 1.7\% and 49$\degr$, respectively.

\begin{figure}
	\includegraphics[height=7cm,width=\columnwidth]{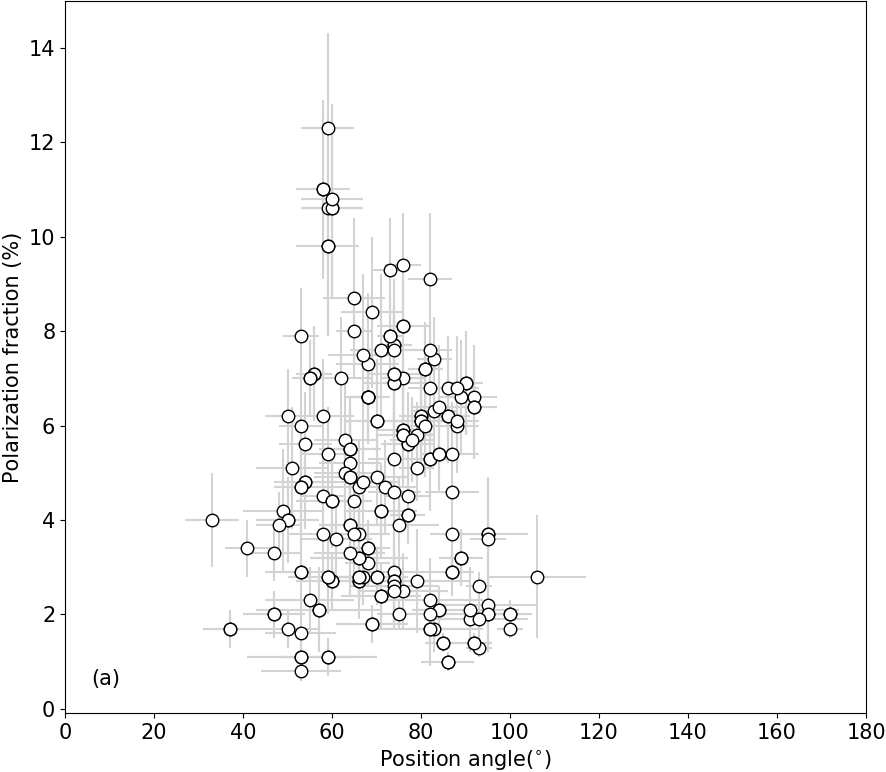}
	\includegraphics[height=7cm,width=\columnwidth]{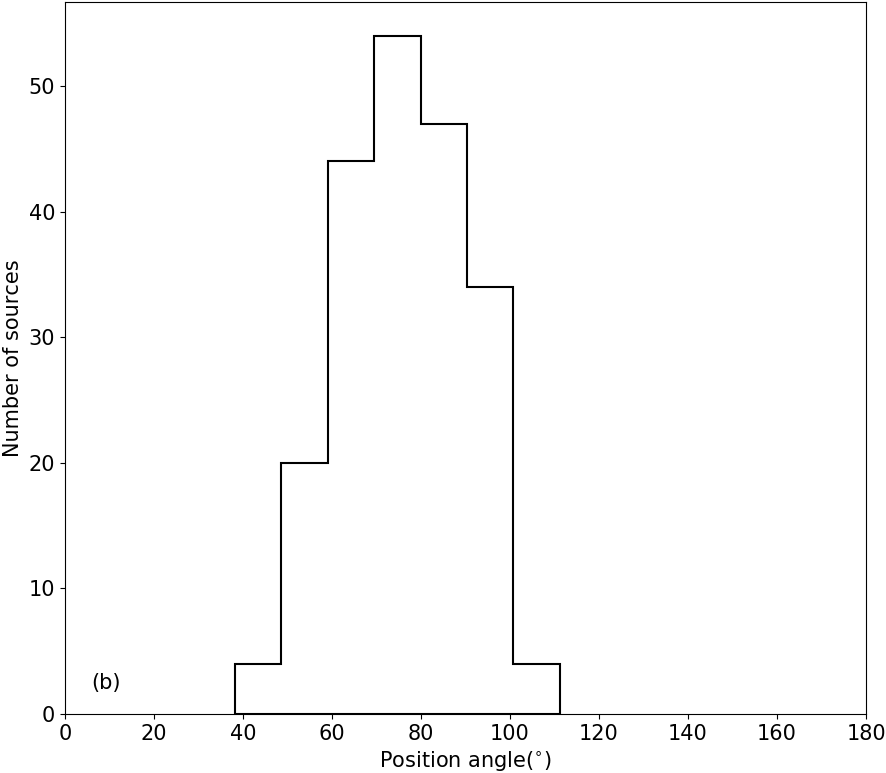}
	\caption{\textbf{(a)} Plot of \pp versus \tp towards L1616 is shown using open circles. \textbf{(b)} Histogram of \tp of the same region is shown with bin size $\sim10\degr$.}
	\label{fig:planck_theta}
\end{figure}

Now, the polarization and position angle of a star depends on its immediate surroundings, e.g., for a young star, a circumstellar dusty disk may contribute to the polarization measurements. Also, the presence of an inhomogeneous density distribution of the cloud could affect the estimated polarization. This effect might be eliminated, and a better picture could be obtained by considering the mean polarization parameters for all the sources within an area of the sky but at the cost of spatial resolution. For this purpose, we divide the area of polarimetric observations into 18$\times$18 grids each of size $5\arcmin\times5\arcmin$. Then we take a mean of $P$ and $\theta$ of all the stars lying within one grid. Fig. \ref{fig:pol_rslts_figure} (b) shows the mean results of each grid using yellow-colored vectors. The grids used in the analysis are also shown using white-colored boxes. From this figure, it is clearly noticeable that the mean directions of polarization vectors follow the cloud structure. A curvature in the \bpos can be noticed in the head region of L1616 and CB 28. Also, the mean polarization vectors are found to show the trend of being twisted along the comma structure of the tail. The range of the mean $P$ and $\theta$ are $0.1-9.1$\% and $12\degr-175\degr$, respectively. The dispersion values of the same are 1.5\% and 38$\degr$, respectively.

The dispersion in $\theta$ can indicate the disordered behavior of \bpos. In Fig. \ref{fig:p_theta_figure} (a), we present the distribution of $P$ with $\theta$ for individual and mean results using open and filled grey circles, respectively. In Fig. \ref{fig:p_theta_figure} (b), we have shown the distribution of $\theta$ values for individual and mean results using open and filled grey histograms, respectively, with a bin size of $\sim10\degr$. The peaks of the histograms before and after taking the mean are $\sim103\degr$ and $\sim113\degr$, respectively. The high dispersion in theta values even after performing the grid-average implies a chaotic nature of the ambient \bpos.

\subsection{Orientation of Magnetic field towards L1616 inferred from \textit{Planck} archival data}

The magnetic field geometry towards L1616 inferred from \textit{Planck} archival data is shown in Fig. \ref{fig:planck_figure}. The \bpos inferred by our R-band measurements is shown using cyan lines with uniform length (3\%). Compared to the optical polarimetric results, \textit{Planck} polarization measurements show a smoother variation in position angles, i.e., the orientation of \bpos\!\!. A bending of position angles is clearly visible in the northern area of the tail part of L1616. Also, \pp is noticeably lower in the head region and other denser parts of the cloud, consistent with the other studies \citep[e.g.,][]{2018ApJ...861...65S, hoang2021studying}.

We have used the threshold of $P_{P}/\Delta{P_{P}}\geqslant2$ in the analysis of \textit{Planck} data. We debias the \textit{Planck} polarization measurements and estimate the uncertainties in \pp and \tp following the equations provided by \cite{2015A&A...574A.135M}. In Fig. \ref{fig:planck_theta} (a), we show the distribution of \pp as a function of \tp. The ranges of the \pp and \tp are $0.8-12.3$\% and $33\degr-106\degr$, respectively. The mean values of \pp and \tp are 4.8\% and $71\degr$, and dispersions of the same are 2.4\% and 14$\degr$, respectively. The distribution of \pp is shown as a form of histogram with bin size $\sim10\degr$ in Fig. \ref{fig:planck_theta} (b). The much lower dispersion of \tp indicates the comparatively regular distribution of \bpos in the sub-millimeter regime. As a whole, \tp is showing an almost perpendicular orientation of \bpos with respect to the direction of the ionizing radiation towards L1616, while the mean $R$-band polarimetric results are showing \bpos following the cloud structure. Because of relatively better resolution, the small scale variation in our observed $R$-band polarimetric results allow us to obtain new information that had not been investigated before. 
 
\subsection{Probable RDI and Rocket Effect ongoing towards L1616}

In order to study the impact of the magnetic field on the molecular clouds, \cite{2009MNRAS.398..157H} investigated the structural transformation of a magnetized globule, photoionized by massive OB stars. Their simulations depicted that the evolution of a photoionized globule can be significantly regulated by the presence of strong magnetic field lines. Even in the absence of magnetic fields, the photoionized globule can evolve in two processes. First is the radiation-driven implosion \citep[RDI;][]{1989ApJ...346..735B}, and the later is the acceleration of the globule via the rocket effect \citep{1955ApJ...121....6O}, creating the elongated structure. During the RDI process, the dense clump gets compressed by shock front until it reaches in equilibrium with the surrounding ionized medium. Due to the rocket effect, the globule accelerates away from the source of ionization, and creates extended structures like elephant trunks, CGs, BRCs, etc., that are usually found at the boundaries of the H{\sc ii} regions \citep[e.g., ][]{1991ApJS...77...59S}. Later, \cite{2011MNRAS.412.2079M} incorporated the magnetic fields of various orientations and strengths (weak, medium, and strong) to the 3-dimensional hydrodynamic simulations, including photoionizing radiative transfer also. They found that due to the RDI process, the weak magnetic field oriented initially at the perpendicular direction, got modified significantly. The affected field lines were altered into alignment with the elongated structure by the dynamics of RDI. For the case of magnetic field lines of medium strength, oriented perpendicularly, the field structure changed marginally. For a strong magnetic field with perpendicular alignment, the change was insignificant.

The RDI mechanism predicts that the cloud would collapse perpendicular to the direction of the ionizing stars. The elongations obtained by \cite{1999PASJ...51..837Y} in the $^{12}$CO, $^{13}$CO, and C$^{18}$O emission contours towards L1616 along the direction of massive Orion stars further supported this mechanism. \cite{2008ApJ...687.1303G} found a small scale sequential age gradient in the PMS stars towards L1616, i.e., younger sources are located to the west of the bright rim, and older sources are found to be distributed eastward where the massive stars reside. The comet-shaped head part of L1616 might have been formed due to the interaction between the ionizing radiation emitted from the OB stars and the cloud through the RDI process. Also, the cluster in the head region might have formed due to the triggering effect of the RDI process. Subsequently, the cloud was evolved dynamically, as an effect of which the initially perpendicular magnetic field lines might get dragged away from the source of ionization following the structure of the globule, which results in the twist of magnetic field lines in the head area of the globule. However, the ionizing radiation might not affect significantly enough to align the magnetic field lines entirely along the direction of ionization, which is found towards a number of regions e.g., CG22 \citep{1996MNRAS.279.1191S}, M16 \citep{2007PASJ...59..507S}, etc., and in the simulations \citep{2009MNRAS.398..157H,2010MNRAS.403..714M, 2011MNRAS.412.2079M}. So, either the initial magnetic field had sufficient strength to resist an entire alignment of the field lines along the direction of ionization, or the distances between the ionizing sources and the globule were higher enough to make the extent of their effect on the globule insignificant. Considering a distance of 384 pc to L1616, the true distance between $\epsilon$ Ori and L1616 is estimated to be $\sim60$ pc, much longer than those adopted in the simulations, where the initial distance between the ionizing source and the globule is less than one pc \citep{2009MNRAS.398..157H,2010MNRAS.403..714M, 2011MNRAS.412.2079M}. 

\cite{1995MNRAS.276..923R} estimated the mass of L1616 (head region) to be 180 M$_{\odot}$. From cloud-averaged line width (2.7 kms$^{-1}$), the corresponding virial mass for keeping the cloud bound was estimated as 1000 M$_{\odot}$, which was about five times higher than the estimated mass. It seems that probably the energy emitted from the massive stars associated with the cluster is disintegrating the cloud. There was also evidence of substantial mass motion from the $^{12}$CO molecular line study described by \cite{1995MNRAS.276..923R}. The excess turbulent motions of $\sim1.5$ kms$^{-1}$ over and above that needed for virial equilibrium and the present size of head of L1616 ($\sim2$ pc) indicate that the disintegration has possibly been occuring since the last $1-2$ Myr \citep{1995MNRAS.276..923R}. In our polarimetric study, the magnetic field geometry is found to be more chaotic in the head region, which might be due to the disintegration of cloud material by the energy input from the reflection nebula and, moreover, the presence of OB stars. The magnetic field follows the cloud's tail region, as can be seen in Fig. \ref{fig:pol_rslts_figure} (a) and (b) both.

\begin{figure*}
	\includegraphics[height=7.4cm, width=16cm]{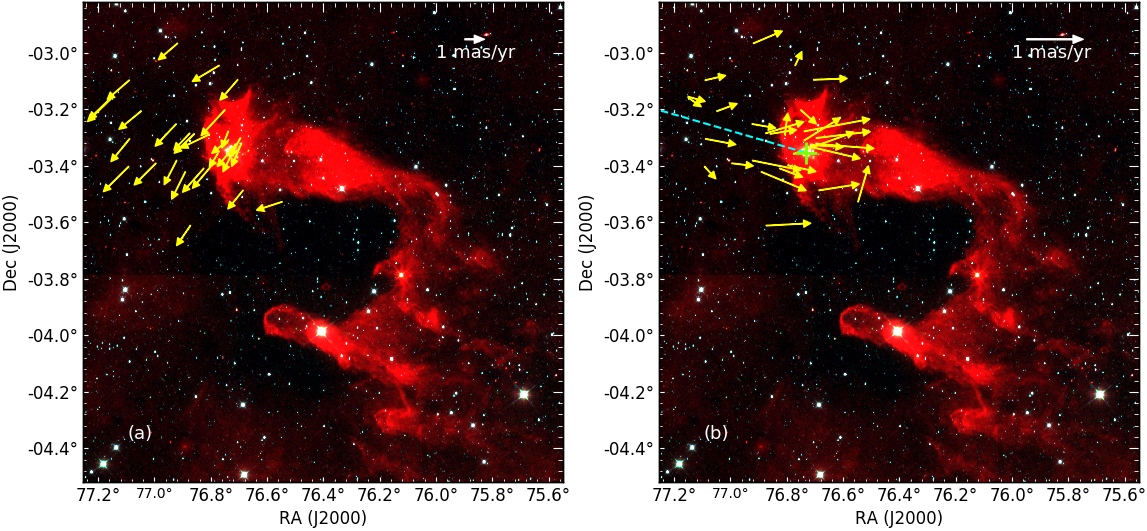}
	\caption{Proper motion directions of the 29 known YSOs in L1616 on the \textit{WISE} color composite diagram using 3.6 (blue), 4.5 (green), and 12 (red) $\mu$m images. \textbf{(a)} The yellow arrows show the proper motion vectors of the YSOs obtained from \textit{Gaia} EDR3. \textbf{(b)} The yellow arrows represent the relative proper motion vectors of the YSOs with respect to the ionizing source $\epsilon$ Ori (not shown in this diagram). The cyan dashed line shows the direction of the ionizing radiation with respect to the \textit{IRAS} source (green `+' symbol) embedded in L1616.}
	\label{fig:YSO_pm_l1616}
\end{figure*} 

As mentioned earlier, in Fig. \ref{fig:search_area} (a), we show the locations of all the O- and B-type stars around L1616, which might influence the cloud by their UV radiation. The distance to these stars is in the range of $\sim323-575$ pc (see Table \ref{tab:ob_stars_table}), most of which have a distance consistent with the distance to L1616. This is to note that RUWE of \#10, i.e., HD 35149 has RUWE=2.697. A higher RUWE makes the astrometric measurements of HD 35149 unreliable. Two stars (\#1 and \#3) for which there are no \textit{Gaia} measurements, the distances have been acquired from \textit{Hipparcos}. Based on the inclination of the tail, the L1616 cloud complex directs to $\epsilon$ Ori \citep[B0Ia;][star \# 3 in Table \ref{tab:ob_stars_table}, also shown in Fig. \ref{fig:search_area} (a)]{1968ApJS...17..371L}, \cite{1995MNRAS.276..923R} suspected this star to be the primary source of ionization and responsible for developing the cometary shape of L1616. He also considered this star responsible for triggering the star formation in it as the line joining the ionizing radiation and peak CO intensity, is directed radially towards $\epsilon$ Ori. However, he mentioned that the influence of the other early-type stars on L1616 can not be neglected. For example, star \# 1, HD 37468 ($\sigma$ Ori, O9.5V). The location of this star is also along the direction of the head of L1616. There is another star \#4, HD 36779, which is of B2.5V spectral type. But as it is of comparatively late-type, the capability of its ionization compared to the other two stars can be considered to be relatively low.

\subsection{Influence of $\epsilon$ Ori and possible Rocket Effect}
 
As the \textit{Gaia} observations are not available for $\epsilon$ Ori, the distance is obtained from parallax information (2.43$\pm$0.91 mas) of the \textit{Hipparcos} \citep{1997A&A...323L..49P}, which converts to a distance of 412$\pm154$ pc. The distance to $\sigma$ Ori of 352$\pm$113 pc is also derived from the parallax measurement (2.84$\pm$0.91 mas) by the \textit{Hipparcos}. We have made an attempt to investigate the pressure components and the extents of the impact of $\epsilon$ Ori and $\sigma$ Ori on L1616 following the method provided by \cite{2019ApJ...885...68B}. The massive stars are considered to affect the surrounding molecular clouds by exerting radiation pressure \citep{2008MNRAS.389.1009S}, ionization pressure \citep{2003ApJ...599.1333S} and wind pressure \citep{2015MNRAS.450.1199D}. The pressure owing to stellar winds ($P_{wind}$) iss estimated using the following equation.
\begin{equation}\label{eq:p_wind}
P_{wind} = \frac{\dot{M}_{W}V_{W}}{4\pi D_{s}^{2}}
\end{equation}
where $\dot{M}_{W}$ is the rate of stellar mass loss, $V_{W}$ is the velocity of the stellar wind and $D_{s}$ denotes the distance from the star. The average mass loss rate and wind velocity are assumed to be around $10^{-6.08}$ M$_{\odot}$ yr$^{-1}$ and 1400 km s$^{-1}$ for $\epsilon$ Ori, respectively \citep{2002MNRAS.337.1309S}. For $\sigma$ Ori, these values are $10^{-6.74}$ M$_{\odot}$ yr$^{-1}$ and 1500 km s$^{-1}$, respctively \citep{2002MNRAS.337.1309S}. 

Now, the pressure driven by stellar radiation ($P_{rad}$) is computed based on the following equation.
\begin{equation}
P_{rad} = \frac{L_{bol}}{4\pi cD_{s}^{2}}
\end{equation}
The bolometric luminosity ($L_{bol}$) of $\epsilon$ Ori and $\sigma$ Ori are estimated using $V$ magnitude with the required bolometric corrections. The absolute magnitudes in $V$-band ($M_{V}$) corresponding to the spectral types of $\epsilon$ Ori and $\sigma$ Ori are obtained from \cite{1973AJ.....78..929P}. Bolometric corrections for the respective spectral types are adopted from \cite{1984ApJ...284..565H}. We consider the typical bolometric magnitude of the sun ($M_{bol\odot}$) as 4.74 \citep{2018A&A...616A...8A} for estimation of $L_{bol}$.

We estimate the ionized gas pressure ($P_{HII}$) using the following equation.
\begin{equation}
P_{HII} = \mu_{II}m_{H}c_{II}^{2}\sqrt{\frac{S_{Lyc}}{4\pi\beta_{2}D_{s}^{3}}}
\end{equation}
where $\mu_{II}$ is the mean molecular weight in an H{\sc ii} region, which is considered to be =0.678 \citep{2009A&A...497..649B}. c$_{II}$ ($=11$ km s$^{-1}$) is the sound speed in an H{\sc ii} region, and $\beta_{2}$ ($=2.6\times$10$^{-13}$ cm$^{3}$ s$^{-1}$) is the recombination coefficient. S$_{Lyc}$ for $\epsilon$ Ori and $\sigma$ Ori are considered as 10$^{48.4}$ and 10$^{47.9}$ photon s$^{-1}$, respectively \citep{2002MNRAS.337.1309S}. Considering the true distances between $\epsilon$ Ori and $\sigma$ Ori and L1616 ($\sim60$ pc, see Table \ref{tab:ob_stars_table}), the total pressure are estimated as $1.28\pm3.46\times10^{-13}$ and $1.96\pm3.46\times10^{-14}$ dynes cm$^{-2}$, respectively. Though the pressure is low at 60 pc, which is the true distance, considering that initially the cloud was at close proximity to the ionizing star, say at 1 pc, the total pressure $P_{tot}(=P_{wind}+P_{rad}+P_{HII}$) of $\epsilon$ Ori and $\sigma$ Ori are estimated as $4.52\times10^{-10}$ and $6.92\times10^{-11}$ dynes cm$^{-2}$, respectively. As in both the distance considerations, the $P_{tot}$ for $\epsilon$ Ori is about one order of magnitude higher than that of $\sigma$ Ori, $\epsilon$ Ori seems to be the main source responsible for influencing L1616. Moreover, based on the rotating stellar tracks, \cite{2010A&A...520A..51V} estimated the ages of $\epsilon$ Ori and $\sigma$ Ori as 5.7 and 3.8 Myr, respectively. Later, \cite{2015ApJ...799..169S} estimated a much younger mean age ($\sim1$ Myr) of the $\sigma$ Ori system. So, $\sigma$ Ori was born relatively later than $\epsilon$ Ori. Thus, the total pressure and age of both the massive stars indicate that L1616 was primarily affected by $\epsilon$ Ori.

The proper motion values, \mua and \mud of $\epsilon$ Ori are taken from the \textit{Hipparcos}, which are 1.49$\pm$0.80 and -1.06$\pm$0.44 mas yr$^{-1}$, respectively. In Fig. \ref{fig:YSO_pm_l1616} (a), we show the observed proper motion vectors of the 29 YSOs using yellow arrows obtained from \textit{Gaia} EDR3. In Fig. \ref{fig:YSO_pm_l1616} (b), the proper motion vectors of the same sources with respect to $\epsilon$ Ori are presented using yellow arrows. The direction of the ionizing radiation is shown by connecting $\epsilon$ Ori with the central \textit{IRAS} (Infrared Astronomical Satellite) source (green `+' symbol) embedded in L1616 using a cyan dashed line. From the figure, it is evident that a majority of the YSOs show relative motion away from the direction of $\epsilon$ Ori, supporting the concept of ``Rocket Effect'' \citep[also see ][]{2022MNRAS.510.2644S}. The relative median proper motion of the YSOs (\mua$=-0.512$ and \mud$=0.024$ mas yr$^{-1}$) yield a tangential velocity $\sim0.933$ km s$^{-1}$, which is close to the typical velocity dispersion of 1 km s$^{-1}$. Fig. \ref{fig:pm_sigma} in appendix shows the relative proper motion vectors of the known YSOs with respect to ther other massive OB stars (listed in table \ref{tab:ob_stars_table}).

\subsection{Velocity distribution towards L1616}

\begin{figure}
	\includegraphics[height=8cm,width=\columnwidth]{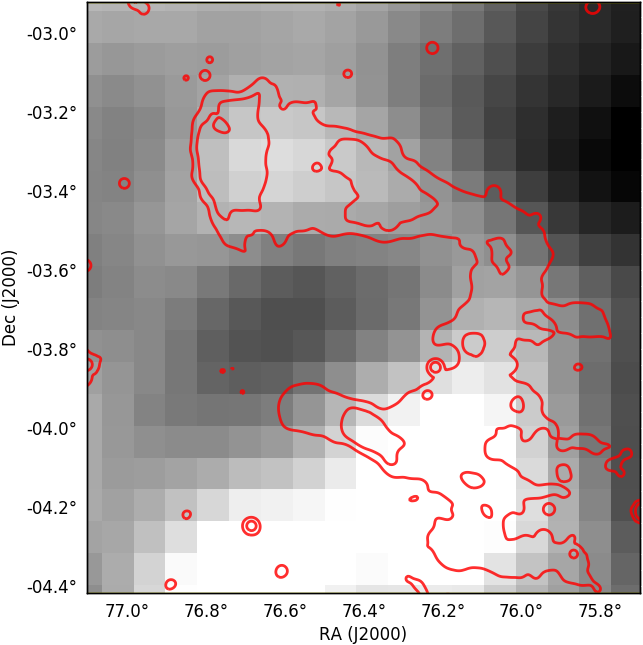}
	\caption{The velocity integrated H{\sc i} image of L1616 and CB28, from HI 4-PI archival data. The contours in red are shown to reveal the cloud structure as can be seen in Fig. \ref{fig:search_area} (b).}
	\label{fig:l1616_hi}
\end{figure}

\cite{1995MNRAS.276..923R} mapped the region towards L1616 in $^{12}$CO and $^{13}$CO ($J=1-0$) molecular lines. The average $^{12}$CO emission from the cloud material was found to be in the ranges $5.4-6.7$ km s$^{-1}$ and $8.7-10.4$ km s$^{-1}$. 
	
We present the H{\sc i} emission towards L1616 in Fig. \ref{fig:l1616_hi} based on the HI 4-PI archival data \citep{2016A&A...594A.116H}. We find that the emission is mainly found towards CB28. As most of the YSOs have been found to be located towards L1616 and none of them towards CB28, it further justifies the absence of H{\sc i} emission in L1616 as most of the cold H{\sc i} gas has been wiped off to form stars. Therefore, the H{\sc i} gas retains towards CB28 and thus shows most of the emission.

\begin{figure}
	\includegraphics[height=9.7cm,width=\columnwidth]{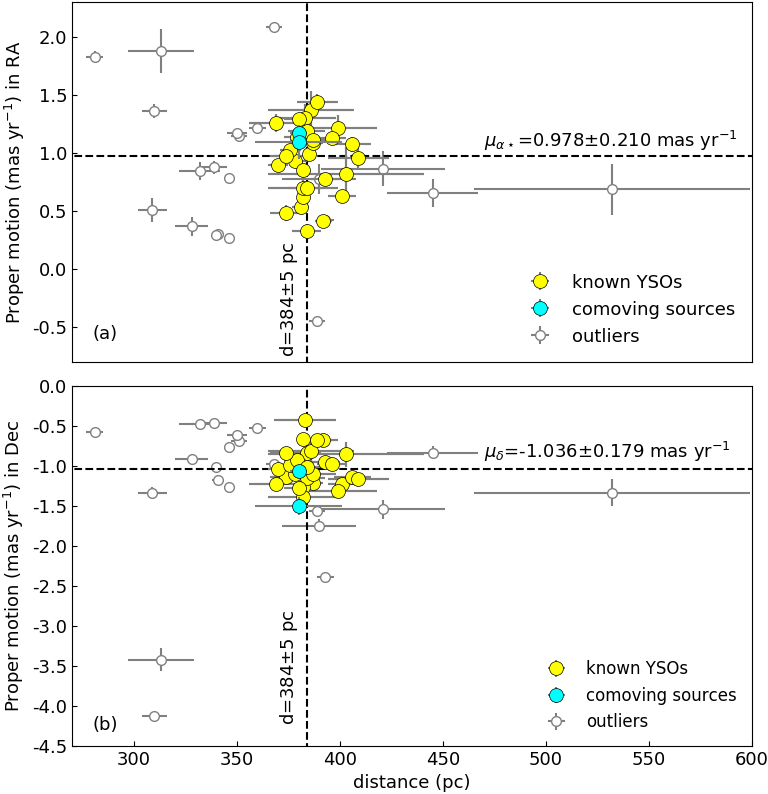}
	\caption{Proper motion versus distance plot of the newly found comoving sources along with the known YSOs. In \textbf{(a)}, the filled yellow circles and the filled cyan circles represent the YSOs and comoving sources for \mua, respectively. In \textbf{(b)}, the symbols represent the same but for \mud. The outliers are shown using open circles. The dashed lines show the median values of $d$, $\mu_{\alpha\star}$, and $\mu_{\delta}$.}
	\label{fig:L1616_comov_dmuamud}
\end{figure}

\begin{figure*}
	\centering
	\includegraphics[height=5.5cm, width=18cm]{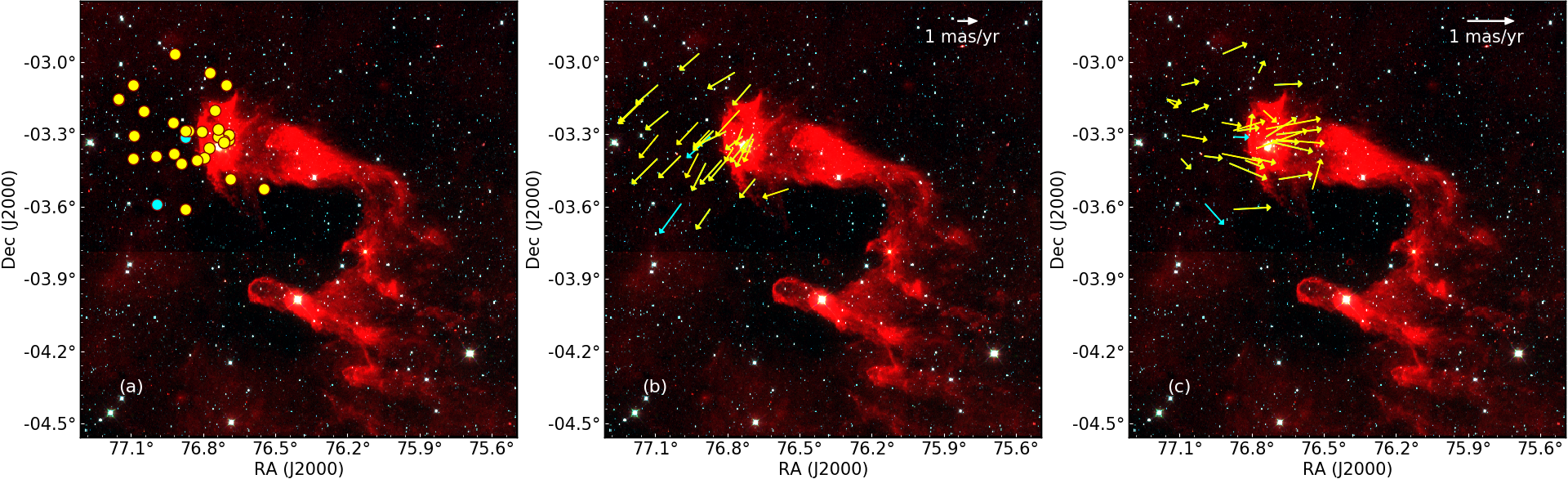}
	\caption{\textbf{(a)} The known YSOs (yellow filled circles) and comoving sources (cyan filled circles) associated with L1616 overplotted on the \textit{WISE} color composite diagram using 3.6 (blue), 4.5 (green), and 12 (red) $\mu$m images. \textbf{(b)} The observed proper motion vectors of the known YSOs (yellow vectors) and the comoving sources (cyan vectors) obtained from \textit{Gaia} EDR3. \textbf{(c)} The relative proper motion vectors of the known YSOs (yellow vectors) and the comoving sources (cyan vectors) with respect to $\epsilon$ Ori.}
	\label{fig:l16_comov}
\end{figure*}

\subsection{Additional comoving sources associated with L1616}

There is a possibility of existence of some young sources which do not have any near-IR excess or significant H$\alpha$ emission, which belong to the WTTS group. Similar to our previous studies \citep{2020MNRAS.494.5851S,2021A&A...653A.142S,2022MNRAS.510.2644S}, we have made an attempt to find these types of sources, based on their distances and proper motions in the same search area. Using the same methodology described in \cite{2020MNRAS.494.5851S}, we obtain two additional comoving sources, which are presented in Table \ref{tab:com_gaia}. The comoving source \#c1, i.e. \textit{Gaia} EDR3 3213013808064290304, is also known as HD 33056. It is of spectral type B9 and mass 2.5M$_{\odot}$ \citep{2005AJ....129..856H}, which belongs to Orion OB1a association. \cite{2004A&A...416..677A} found X-ray emission associated with this star based on the RASS observations. However, no H$\alpha$ emission was found in its spectra \citep{2005AJ....129..856H}. Recently, \cite{2019A&A...623A..72K} detected a possible physical companion orbiting HD 33056 based on its proper motion anomaly. In Fig. \ref{fig:L1616_comov_dmuamud}, we have shown the distribution of the comoving sources in the distance versus proper motion plane, along with the previously known YSOs. We have shown the spatial distribution of those comoving sources using filled cyan circles in Fig. \ref{fig:l16_comov} (a). From this figure, it is clear that only L1616 harbors a majority of the YSOs and additional comoving sources (mostly in the head region and in towards the north-east direction, which is the direction of the massive ionized sources also), while CB28 does not. This is further supported by the fact that only the region close to CB28 shows higher H{\sc i} emission, while L1616 is deprived of it because already star formation has been taken place in this cloud, which is evident from Fig. \ref{fig:l1616_hi}. The observed proper motions of these two types of sources are shown in Fig. \ref{fig:l16_comov} (b), using yellow and cyan-colored vectors. The differential proper motions of all the sources with respect to $\epsilon$ Ori are also shown in Fig. \ref{fig:l16_comov} (c), using the same symbols. After the selection of the comoving sources based on their distances and proper motions, we investigate the nature of these sources, which are described in the following sections.
 
\begin{figure}
	\includegraphics[width=\columnwidth]{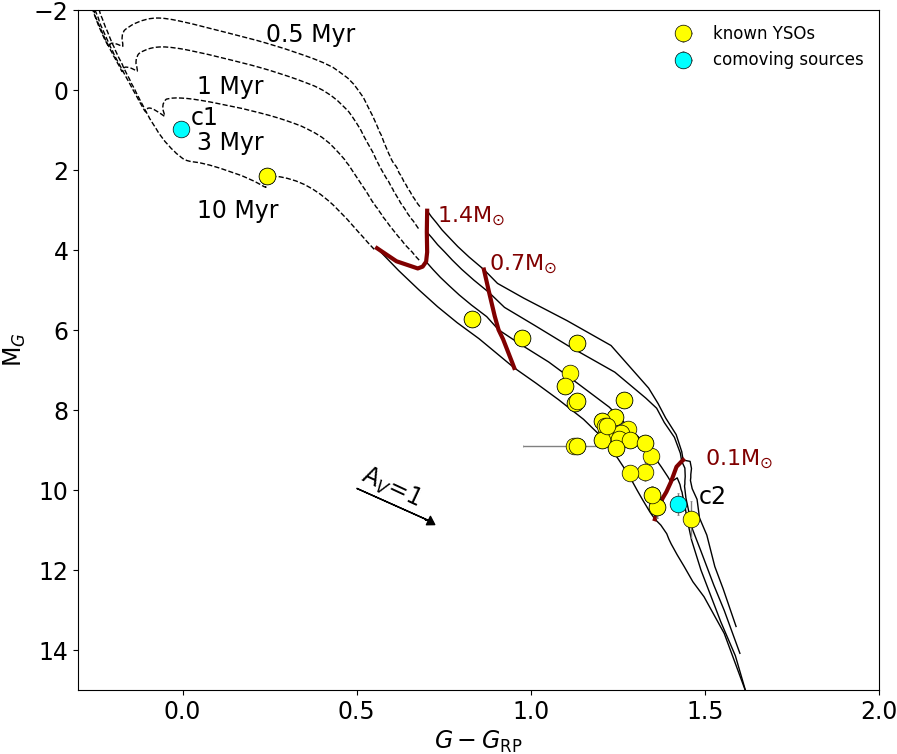}
	\caption{The M$_{G}$ versus ($G-G_\mathrm{RP}$) color-magnitude diagram of the known YSOs (filled circles in yellow) and the comoving sources (filled circles in cyan). The dashed and solid curves in black represent the isochrones from PARSEC models \citep{2017ApJ...835...77M} and CIFIST models \citep{2015A&A...577A..42B}, respectively, for 0.5, 1, 3, and 10 Myr. The stellar tracks for 0.1, 0.7, and 1.4 M$_{\odot}$ are shown using thick brown curves.}
	\label{fig:gaia_yso_comov}
\end{figure}

\subsection{Properties of the comoving sources in L1616}

\subsubsection{The \textit{Gaia} color-magnitude diagram}

The \textit{Gaia} M$_{G}$ versus ($G-G_\mathrm{RP}$) color-magnitude diagram (CMD) for the previously known YSOs and the newly found comoving sources is shown in Fig. \ref{fig:gaia_yso_comov} using yellow and cyan filled circles, respectively. The PMS isochrones corresponding to 0.5, 1, 3, and 10 Myr are also shown in Fig. \ref{fig:gaia_yso_comov}. The dashed curves in black are the PADOVA tracks Parsec 3.5\footnote{\url{http://stev.oapd.inaf.it/cgi-bin/cmd}}\citep{2017ApJ...835...77M} for the higher-mass stars and the thick curves in black are the CIFIST 2011\_2015\footnote{\url{phoenix.ens-lyon.fr/Grids/BT-Settl/CIFIST2011\_2015/ISOCHRONES/}}\citep{2015A&A...577A..42B} models for low-mass stars. Among the 29 YSOs, eight of them were identified by both \cite{2004A&A...416..677A} and \cite{2008ApJ...687.1303G}, and 10 of them were detected by \cite{2008ApJ...687.1303G}. As we do not have information of A$_{V}$ for almost half of the sources, the M$_{G}$ versus ($G-G_\mathrm{RP}$) CMD has been shown without extinction correction. From Fig. \ref{fig:gaia_yso_comov}, it is well evident that most of the sources located towards L1616 are of age $3-10$ Myr. This age distribution is consistent with the same of $\epsilon$ Ori, while $\sigma$ Ori is found to be relatively younger than the YSOs, indicating it might not be the main source to influence L1616. In the study of pre-main sequence stars, \cite{2004A&A...416..677A} and \cite{2008ApJ...687.1303G} estimated their A$_{V}$, most of which are below 1.0. However, a number of them showed a higher A$_{V}$, indicating the possibility of being surrounded by a thick circumstellar disk or highly embedded in their parent cloud.

\begin{figure}
	\includegraphics[width=\columnwidth]{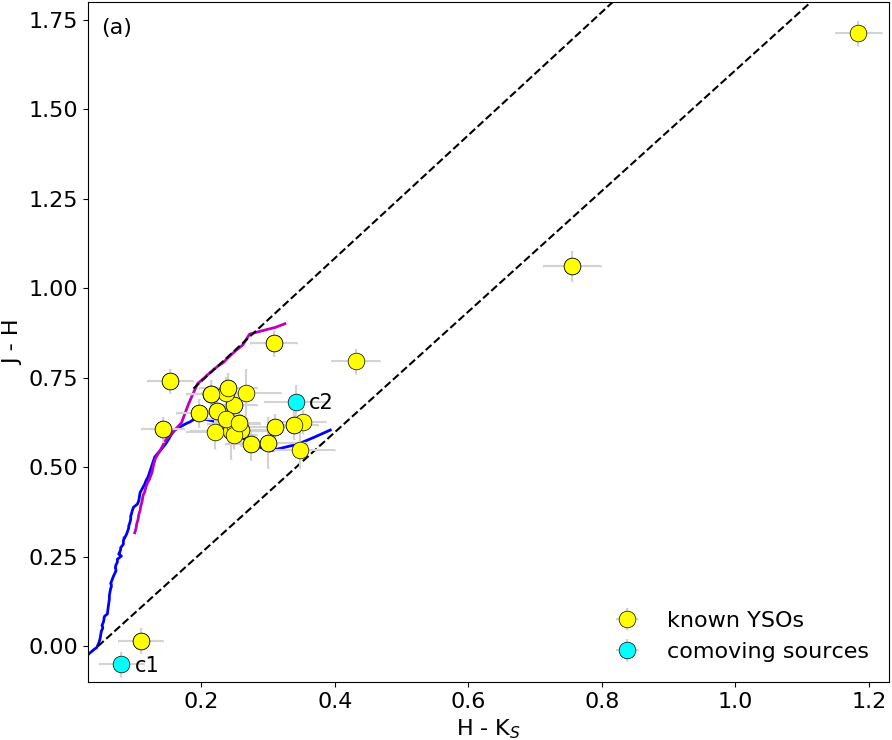}
	\includegraphics[width=\columnwidth]{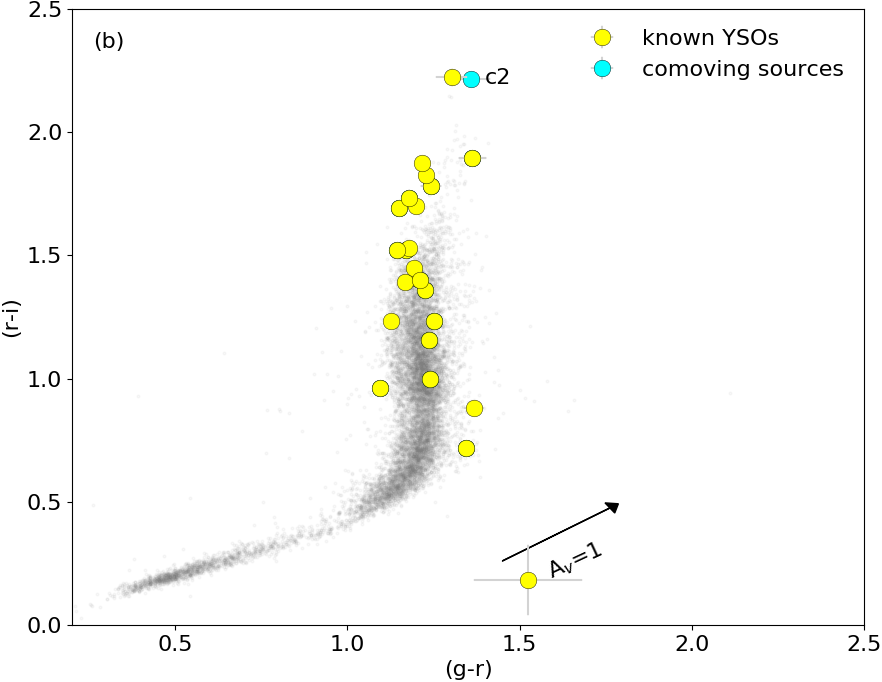}
	\caption{\textbf{(a)} The 2MASS $(J-H)$ versus $(H-K_{S})$ CC diagram for the known YSOs (filled circles in yellow) and the newly found comoving sources (filled circles in cyan). The solid curves in magenta and blue represent the loci of the unreddened giants and main sequence stars, respectively. The dashed lines show the direction of reddening vectors. Two sources, with $H-K_{s}\sim0.75$ and $\sim1.2$, are CTTS named as TTS $050713.5-031722$ and TTS $050646.1-031922$, respectively. \textbf{(b)} The PanSTARRS $(r-i)$ versus $(g-r)$ CC diagram for the same sources (Symbols indicate the same as \textbf{(a)}). The unreddened M-type sources from literature are shown using grey points. The arrow represents the reddening vector of A$_{V}=1$ magnitude.}
	\label{fig:2mass_ps_yso_comov}
\end{figure}

\subsubsection{The $(J-H)$ versus $(H-K_{S})$ color-color diagram}

Fig. \ref{fig:2mass_ps_yso_comov} (a) shows the 2MASS (Two Micron All Sky Survey) $(J-H)$ versus $(H-K_{S})$ color-color (CC) diagram of the known YSOs and the comoving sources. The $J$, $H$, and $K_{S}$ magnitudes of these sources are adopted from \cite{2006AJ....131.1163S}, using a search radius of 1\arcsec. A majority of the sources in this area do not have good quality \textit{WISE} magnitudes. The 2MASS counterparts for the two comoving sources are shown using filled cyan circle and labelled. Of the 29 known YSOs having reliable \textit{Gaia} EDR3 data, 2MASS data for 28 of them are found, which are shown using filled yellow circles in Fig. \ref{fig:2mass_ps_yso_comov} (a).

In Fig. \ref{fig:2mass_ps_yso_comov} (a), a majority of the YSOs and comoving sources show relatively less amount of extinction and near-IR excess. The results of lower A$_{V}$ for most of the YSOs, as obtained by \cite{2004A&A...416..677A} and \cite{2008ApJ...687.1303G}, are consistent with Fig. \ref{fig:2mass_ps_yso_comov} (a). Also, most of the sources found in the vicinity of L1616 are located in the area occupied by M-dwarfs. As a majority of the sources are relatively older ($3-10$ Myr) to sustain circumstellar disks, it can be expected that the sources would suffer less near-IR excess and less extinction. The locations of the sources in Fig. \ref{fig:2mass_ps_yso_comov} (a) agree well with this. 

However, two YSOs in this plot show distinct positions, in terms of relatively higher $(J-H)$ and $(H-K_{S})$ colors. The YSO having $J-H\sim1.7$ and $H-K_{s}\sim1.2$ is star \#3 in Table \ref{tab:YSO_gaia}. This is catalogued as TTS $050646.1-031922$ in both \cite{2004A&A...416..677A} and \cite{2008ApJ...687.1303G}. It is a CTTS, with a higher H$\alpha$ emission ($-41.00\pm2.00$ \AA) and possessing higher A$_{V}$ ($4.33\pm0.35$ mag), obtained by \cite{2008ApJ...687.1303G}. Another YSO, having $J-H\sim1.0$ and $H-K_{s}\sim0.75$, is listed as star \#14 in Table \ref{tab:YSO_gaia}. This is catalogued as TTS $050713.5-031722$ in \cite{2008ApJ...687.1303G}. It is also classified as a CTTS, showing a higher H$\alpha$ emission ($-10.50\pm0.50$ \AA) and higher A$_{V}$ ($1.18\pm0.15$ mag), found by \cite{2008ApJ...687.1303G}.

\subsubsection{The $(g-r)$ versus $(r-i)$ color-color diagram}

The $(g-r)$ versus $(r-i)$ CC diagram for the known YSOs and the comoving sources is presented in Fig. \ref{fig:2mass_ps_yso_comov} (b). The $g$, $r$, and $i$ magnitudes have been retrieved from the Pan-STARRs catalog \citep{2016arXiv161205560C}, by giving a search radius of 1$\arcsec$ for every source. Out of 29 previously known YSOs, the $g$, $r$, and $i$ data are obtained for 24 of them, which are shown using yellow filled circles. Out of two comoving sources, we have acquired $g$, $r$, and $i$ data for one source (\#c2), which is shown using filled circle in cyan. The grey dots represent the population of main-sequence stars selected from an area around the star 10 Lac having spectral types in the range of A0 $-$ M7 \citep{2021A&A...653A.142S}. The spectral types of the sources are obtained from the Simbad database. The arrow in Fig. \ref{fig:2mass_ps_yso_comov} presents the reddening vector of A$_{V}=1$ magnitude.

The population of unreddened M-dwarfs follow a very unique almost `S'-shaped locus in the $(g-r)$ versus $(r-i)$ CC diagram, which help us to clearly distinguish them from the giants and early main-sequence stars. From Fig. \ref{fig:2mass_ps_yso_comov} (b) it is evident that the previously known YSOs and comoving sources are predominantly M-dwarfs, and they show a negligible foreground extinction, as they populate the unreddeded M-dwarf locus, consistent with the conclusion found in Fig. \ref{fig:2mass_ps_yso_comov} (a) and the lower A$_{V}$ estimated by \cite{2004A&A...416..677A} and \cite{2008ApJ...687.1303G} for a majority of the pre-main sequence stars in L1616.

\section{Summary and Conclusions}\label{sec:con}

The first optical polarimetric study to investigate the ambient \bpos geometry towards the L1616 cloud complex is reported in this work. We compare our results with \textit{Planck} archival data. Based on the recently released \textit{Gaia} EDR3 measurements of distances of the associated YSOs, we independently determine the distance to L1616, which is found to be consistent with the previously estimated values. An active RDI mechanism towards L1616 is noted possibly by $\epsilon$ Ori. We also obtain a number of comoving sources having similar distance and proper motions of the previously known YSOs. The main outcomes of this study are summarized below:

\begin{itemize}
	
\item Using the \gaia EDR3 measurements, we obtain the distance to L1616 as 384$\pm$5 pc. The median values of \mua and \mud of the YSOs are estimated to be 0.978$\pm$0.210 and \mud=-1.036$\pm$0.179 mas yr$^{-1}$, respectively. The YSOs show a trend of rocket effect with respect to $\epsilon$ Ori.	

\item The \bpos estimated from $R$-band imaging polarimetric data is found to be random while the measurement of each individual source is considered. However, their mean Bpos values for a $5\arcmin\times5\arcmin$ gird follow the large-scale cloud structure.

\item The \bpos obtained from \planck observations is found to be relatively much smooth compared to the optical $R$-band results, and appears to be almost perpendicular to the direction of ionization. The poor angular resolution of \planck possibly could not reveal the small scale variation of \bpos.

\item The alignment of mean \bpos indicates a possible scenario, where the initial direction of the magnetic field lines was perpendicular to the direction of the ionizing radiation, prior to being affected by $\epsilon$ Ori. Based on our $R$-band polarimetric observations, the magnetic field lines towards L1616 are found to follow the curvature of the head of the cloud complex, which could possibly be because the magnetic field lines might have been pulled away by the strong ionizing radiation.

\item Based on the distance and proper motion measurements of the YSOs, two additional comoving sources are found towards L1616. One (\#c1) of the two newly found comoving sources is HD 33056, a B9 star.

\item Based on the \textit{Gaia} color-magnitude diagram, a majority of the YSOs located towards L1616 are of age $3-10$ Myr. 

\item The Pan-STARR and 2MASS CC diagrams reveal that the sources are predominantly of low-mass M types and suffer from little or no near-IR excess and extinction. Although, the presence of a number of YSOs with higher extinction and higher IR excess can be seen.

\end{itemize}

In order to investigate thoroughly the variation of the magnetic field and how it is related to the evolution of L1616, we need more observations towards this region using molecular lines with better resolution. The small-scale changes in the velocities of different molecular species will enable us to understand the present dynamical state of the cloud in more detail and the RDI process as well. Therefore, the molecular line observations towards L1616 are planned in the near future.

\section*{Acknowledgements}

We sincerely thank the reviewer, Dr. Davide Gandolfi, for his insightful and constructive remarks which significantly improved the manuscript. We acknowledge all the supporting staff at ARIES, Nainital, who made the polarimetric observations possible. We also thank the support by the S. N. Bose National Centre for Basic Sciences under the Department of Science and Technology, Govt. of India. This research has made use of the SIMBAD database, operated at CDS, Strasbourg, France. We also acknowledge the use of NASA's SkyView facility {\url{http://skyview.gsfc.nasa.gov}} located at the NASA Goddard Space Flight Center. This research has made use of the NASA/ IPAC Infrared Science Archive, which is operated by the Jet Propulsion Laboratory, California Institute of Technology, under contract with the National Aeronautics and Space Administration. \textit{Planck} data have been acquired from {\url{http://irsa.ipac.caltech.edu}} or {\url{http://irsa.ipac.caltech.edu/applications/planck/}}. This work has made use of data from the European Space Agency (ESA) mission \textit{Gaia} (\url{https://www.cosmos.esa.int/gaia}), processed by the \textit{Gaia} Data Processing and Analysis Consortium (DPAC, \url{https://www.cosmos.esa.int/web/gaia/dpac/consortium}). Funding for the DPAC has been provided by national institutions, in particular the institutions participating in the \textit{Gaia} Multilateral Agreement. We made use of the Two Micron All Sky Survey, which is a joint project of the University of Massachusetts and the Infrared Processing and Analysis Center/California Institute of Technology, funded by the National Aeronautics and Space Administration and the National Science Foundation. The Pan-STARRS1 Surveys (PS1) have been made possible through contributions of the Institute for Astronomy, the University of Hawaii, the Pan-STARRS Project Office, the Max-Planck Society and its participating institutes, the Max Planck Institute for Astronomy, Heidelberg and the Max Planck Institute for Extraterrestrial Physics, Garching, The Johns Hopkins University, Durham University, the University of Edinburgh, Queen's University Belfast, the Harvard-Smithsonian Center for Astrophysics, the Las Cumbres Observatory Global Telescope Network Incorporated, the National Central University of Taiwan, the Space Telescope Science Institute, the National Aeronautics and Space Administration under Grant No. NNX08AR22G issued through the Planetary Science Division of the NASA Science Mission Directorate, the National Science Foundation under Grant No. AST-1238877, the University of Maryland, and Eotvos Lorand University (ELTE).

\section*{Data Availability}

The $R$-band polarimetric data presented in this work are available in this article. The \textit{Gaia} proper motions and distances of the stars are available in \url{https://vizier.u-strasbg.fr/viz-bin/VizieR-3?-source=I/350/gaiaedr3} and \url{https://vizier.u-strasbg.fr/viz-bin/VizieR?-source=I/352}, respectively. The HI4PI data can be otained from \url{https://cdsarc.cds.unistra.fr/viz-bin/cat/J/A+A/594/A116}. The 2MASS and Pan-STARRS data used in this article are available in \url{https://vizier.u-strasbg.fr/viz-bin/VizieR?-source=II/246} and \url{https://vizier.u-strasbg.fr/viz-bin/VizieR?-source=II/349}, respectively.


\bibliographystyle{mnras}
\bibliography{ref} 



\appendix

\section{Tables}
\begin{table*}
	\begin{center}
		\caption{Properties of the known YSOs identified towards L1616 from \textit{Gaia} EDR3.}
		\label{tab:YSO_gaia}
		\renewcommand{\arraystretch}{1.3}
		\footnotesize
		\begin{tabular}{lccccccccc} 
			\hline
			\#& RA(2016)  & Dec(2016)  & Source ID& $d(\Delta d)$ & $\mu_{\alpha\star}$ ($\Delta\mu_{\alpha\star}$) & $\mu_{\delta}$ ($\Delta\mu_{\delta}$) & $G(\Delta G)$ & $G_\mathrm{BP}(\Delta G_\mathrm{BP})$ & $G_\mathrm{RP}(\Delta G_\mathrm{RP})$\\
			& ($^{\circ}$) & ($^{\circ}$) & (\textit{Gaia} EDR3) & (pc)& (mas yr$^{-1}$) & (mas yr$^{-1}$) & (mag) & (mag) & (mag)\\
			(1)&(2)&(3)&(4)&(5)&(6)&(7)&(8)&(9)&(10)\\
			\hline
1&76.546639&-3.527516& 3212979620124674944 &383$^{13}_{-16}$
&1.305$\pm$0.114&-0.426$\pm$0.080&17.078$\pm$0.003&19.118$\pm$0.030& 15.732$\pm$0.005\\ 
2&76.685127&-3.486913& 3213002744228578048 &393$^{8}_{-7}$ &0.776$\pm$0.053&-0.948$\pm$0.039&16.559$\pm$0.003&18.141$\pm$0.012& 15.344$\pm$0.006\\ 
3&76.691909&-3.322872& 3213015216813585792 &374$^{9}_{-8}$ &0.482$\pm$0.066&-1.138$\pm$0.049&16.764$\pm$0.048& 17.749$\pm$0.123& 15.641$\pm$0.138\\ 
4&76.694368&-3.301577& 3213016007087564160&381$^{4}_{-4}$ &0.535$\pm$0.024&-0.946$\pm$0.017&14.106$\pm$0.003&15.033$\pm$0.004& 13.132$\pm$0.004\\ 
5&76.704082&-3.095333& 3213032293603522432 &370$^{5}_{-6}$ &0.896$\pm$0.043&-1.036$\pm$0.032&16.018$\pm$0.003&17.674$\pm$0.008& 14.775$\pm$0.004\\ 
6&76.711790&-3.326377& 3213015148092614656 &378$^{6}_{-6}$ &0.928$\pm$0.053&-1.096$\pm$0.040&15.628$\pm$0.003&16.748$\pm$0.012& 14.362$\pm$0.005\\ 
7&76.712735&-3.333325& 3213015148094107392 &382$^{2}_{-3}$ &0.624$\pm$0.018&-1.264$\pm$0.013&10.051$\pm$0.003&10.170$\pm$0.003 & 9.810$\pm$0.004\\ 
8&76.737263&-3.309870& 3213015319892792960 &382$^{4}_{-3}$ 
&0.850$\pm$0.022&-0.660$\pm$0.015&14.230$\pm$0.003&15.451$\pm$0.007& 13.098$\pm$0.005\\ 
9&76.737380&-3.277945& 3213015766569389568 &384$^{7}_{-6}$ &0.328$\pm$0.053&-0.842$\pm$0.039&16.395$\pm$0.004&17.996$\pm$0.034& 15.117$\pm$0.008\\ 
10&76.749492&-3.202306& 3213018545412381824 &399$^{21}_{-17}$ &1.215$\pm$0.109&-1.307$\pm$0.075&17.549$\pm$0.003&19.636$\pm$0.031& 16.220$\pm$0.005\\ 
11&76.769670&-3.044740& 3214533706794971136 &386$^{22}_{-20}$ 
&1.371$\pm$0.164&-0.812$\pm$0.122&18.369$\pm$0.003&20.393$\pm$0.067& 17.007$\pm$0.007\\ 
12&76.775401&-3.357854& 3213011780839743360 &392$^{4}_{-3}$ &0.418$\pm$0.028&-0.675$\pm$0.019&15.047$\pm$0.003&16.196$\pm$0.012& 13.934$\pm$0.008\\ 
13&76.795606&-3.398180& 3213010750047595520 &385$^{2}_{-3}$ &0.993$\pm$0.016&-1.171$\pm$0.011&13.664$\pm$0.003&14.383$\pm$0.007& 12.832$\pm$0.006\\ 
14&76.806294&-3.289596& 3213015594770689664 &389$^{11}_{-9}$ &1.438$\pm$0.073&-0.681$\pm$0.053&16.838$\pm$0.003&17.843$\pm$0.010& 15.705$\pm$0.007\\ 
15&76.824373&-3.409185& 3213010509529424384 &387$^{6}_{-4}$ &1.085$\pm$0.037&-1.215$\pm$0.027&15.768$\pm$0.004&17.039$\pm$0.010& 14.641$\pm$0.007\\ 
16&76.858017&-3.286800& 3213013876783764352 &376$^{6}_{-4}$ 
&1.025$\pm$0.045&-0.988$\pm$0.032&16.152$\pm$0.003&17.622$\pm$0.010& 14.947$\pm$0.006\\ 
17&76.872020&-3.611690& 3212951479498928128 &384$^{8}_{-10}$ 
&0.702$\pm$0.059&-1.018$\pm$0.046&16.753$\pm$0.006&18.683$\pm$0.027& 15.424$\pm$0.017\\ 
18&76.874111&-3.284813& 3213013975567071488 &403$^{38}_{-38}$ &0.815$\pm$0.237&-0.856$\pm$0.160&18.758$\pm$0.003&20.707$\pm$0.137& 17.297$\pm$0.009\\ 
19&76.889879&-3.421316& 3212963810348859392 &382$^{19}_{-14}$ &0.696$\pm$0.148&-1.391$\pm$0.109&18.032$\pm$0.003&19.934$\pm$0.034& 16.683$\pm$0.008\\ 
20&76.917441&-2.966262& 3214536081913183104 &374$^{8}_{-8}$ &0.978$\pm$0.061&-0.842$\pm$0.042&16.608$\pm$0.003&18.132$\pm$0.010& 15.404$\pm$0.004\\ 
21&76.920860&-3.381351& 3212964192602089984 &401$^{8}_{-7}$ &0.632$\pm$0.052&-1.230$\pm$0.038&16.426$\pm$0.003&17.968$\pm$0.007& 15.214$\pm$0.004\\ 
22&76.922273&-3.251851& 3213014460899305088 &406$^{9}_{-9}$ &1.073$\pm$0.066&-1.132$\pm$0.044&16.609$\pm$0.003&18.328$\pm$0.012& 15.350$\pm$0.004\\ 
23&76.994275&-3.390995& 3212964707998158720 &387$^{12}_{-9}$ &1.110$\pm$0.063&-1.106$\pm$0.048&16.881$\pm$0.003&18.511$\pm$0.017& 15.638$\pm$0.005\\ 
24&77.046349&-3.206081& 3214515362990965376 &379$^{7}_{-6}$ &1.135$\pm$0.042&-0.928$\pm$0.029&15.669$\pm$0.003&16.966$\pm$0.006& 14.537$\pm$0.005\\ 
25&77.087562&-3.304446& 3214466670945275008 &409$^{15}_{-15}$ 
&0.953$\pm$0.103&-1.160$\pm$0.078&17.637$\pm$0.003&19.471$\pm$0.033& 16.352$\pm$0.005\\ 
26&77.088328&-3.096577& 3214518661525826688 &396$^{8}_{-6}$ &1.128$\pm$0.035&-0.977$\pm$0.023&15.396$\pm$0.003&16.573$\pm$0.010& 14.298$\pm$0.006\\ 
27&77.090974&-3.402553& 3212961753060648960 &380$^{7}_{-7}$ &1.291$\pm$0.052&-1.277$\pm$0.040&16.288$\pm$0.003&17.824$\pm$0.008& 15.070$\pm$0.004\\ 
28&77.149065&-3.158139& 3214470454814173568 &369$^{14}_{-12}$ &1.259$\pm$0.079&-1.222$\pm$0.059&16.559$\pm$0.003&18.098$\pm$0.015& 15.307$\pm$0.005\\ 
29&77.149230&-3.154988& 3214470557892112000 &384$^{9}_{-9}$
&1.188$\pm$0.066&-1.144$\pm$0.049&16.683$\pm$0.003&18.503$\pm$0.018& 15.400$\pm$0.005\\ 
    \hline
\end{tabular}\\
\end{center}
\end{table*}
\begin{table}
	\centering
	\caption{Polarization results (debiased) of observed stars located at the background of L1616.}
	\label{tab:pol_rslt_table}
	\begin{tabular}{lcccr} 
		\hline
		Star &  RA(2016) & Dec(2016) & $P\pm\Delta P$ & $\theta\pm\Delta\theta$ \\
		ID  & ($^{\circ}$) & ($^{\circ}$) & (\%) & ($^{\circ}$)\\
		(1) & (2) & (3) & (4) & (5)\\	
		\hline
1 & 75.891327 & -3.576771 & 0.9$\pm$0.3 & 135$\pm$9\\ 
2 & 75.906107 & -3.567638 & 0.4$\pm$0.2 & 150$\pm$10\\ 
3 & 75.932119 & -3.588007 & 1.3$\pm$0.3 & 156$\pm$8\\ 
4 & 75.946613 & -3.622799 & 4.1$\pm$0.1 & 12$\pm$5\\ 
5 & 75.946664 & -3.529529 & 0.7$\pm$0.1 & 152$\pm$5\\ 
6 & 75.958663 & -3.630865 & 3.0$\pm$0.6 & 103$\pm$6\\ 
7 & 75.982440 & -3.565650 & 0.9$\pm$0.1 & 144$\pm$5\\ 
8 & 76.014303 & -3.636689 & 1.0$\pm$0.1 & 152$\pm$5\\ 
9 & 76.024144 & -3.626344 & 0.6$\pm$0.1 & 150$\pm$5\\ 
10 & 76.026911 & -3.647879 & 0.7$\pm$0.2 & 144$\pm$6\\ 
11 & 76.080022 & -3.699110 & 1.4$\pm$0.2 & 161$\pm$4\\ 
12 & 76.083973 & -3.651224 & 1.0$\pm$0.3 & 118$\pm$9\\ 
13 & 76.087104 & -3.667291 & 0.7$\pm$0.1 & 156$\pm$8\\ 
14 & 76.091244 & -3.459438 & 1.0$\pm$0.4 & 90$\pm$11\\ 
15 & 76.106401 & -3.556611 & 7.7$\pm$0.8 & 16$\pm$3\\ 
16 & 76.116816 & -3.525667 & 1.6$\pm$0.6 & 45$\pm$12\\ 
17 & 76.128743 & -3.464229 & 0.2$\pm$0.1 & 65$\pm$8\\ 
18 & 76.131367 & -3.469218 & 0.7$\pm$0.3 & 66$\pm$12\\ 
19 & 76.133008 & -3.521975 & 2.1$\pm$0.4 & 23$\pm$6\\ 
20 & 76.133981 & -3.473007 & 0.6$\pm$0.2 & 78$\pm$7\\ 
21 & 76.177981 & -3.436289 & 2.4$\pm$1.4 & 0$\pm$16\\ 
22 & 76.208430 & -3.975661 & 0.3$\pm$0.1 & 106$\pm$8\\ 
23 & 76.223054 & -4.122173 & 0.8$\pm$0.4 & 63$\pm$12\\ 
24 & 76.229715 & -3.432872 & 0.6$\pm$0.1 & 141$\pm$5\\ 
25 & 76.233140 & -3.960366 & 2.2$\pm$0.4 & 116$\pm$5\\ 
26 & 76.238237 & -3.446391 & 1.6$\pm$0.9 & 120$\pm$14\\  
27 & 76.238237 & -3.988340 & 1.7$\pm$0.2 & 66$\pm$20\\ 
28 & 76.247971 & -3.988116 & 0.2$\pm$0.1 & 73$\pm$11\\ 
29 & 76.248707 & -4.005096 & 3.7$\pm$1.9 & 60$\pm$26\\ 
30 & 76.249795 & -3.340905 & 4.2$\pm$0.9 & 75$\pm$7\\ 
31 & 76.256180 & -4.091339 & 2.0$\pm$0.2 & 25$\pm$16\\ 
32 & 76.262285 & -3.932249 & 0.5$\pm$0.0 & 116$\pm$13\\ 
33 & 76.269078 & -4.093950 & 5.1$\pm$1.1 & 149$\pm$6\\ 
34 & 76.269471 & -4.085907 & 2.1$\pm$0.6 & 119$\pm$20\\ 
35 & 76.271555 & -3.358235 & 1.1$\pm$0.5 & 62$\pm$12\\ 
36 & 76.276157 & -3.464332 & 2.7$\pm$1.6 & 107$\pm$15\\ 
37 & 76.278469 & -4.010696 & 0.7$\pm$0.4 & 75$\pm$13\\ 
38 & 76.279461 & -4.142948 & 0.5$\pm$0.2 & 162$\pm$9\\ 
39 & 76.282942 & -3.346758 & 2.3$\pm$1.0 & 24$\pm$14\\ 
40 & 76.286526 & -4.114692 & 0.7$\pm$0.3 & 9$\pm$13\\ 
41 & 76.291672 & -4.067818 & 1.4$\pm$0.3 & 69$\pm$14\\ 
42 & 76.294342 & -4.001120 & 0.7$\pm$0.2 & 123$\pm$3\\ 
43 & 76.307837 & -4.082435 & 0.5$\pm$0.1 & 4$\pm$5\\ 
44 & 76.309448 & -4.166831 & 0.2$\pm$0.1 & 125$\pm$8\\ 
45 & 76.321199 & -3.360789 & 3.5$\pm$0.9 & 79$\pm$8\\ 
46 & 76.323484 & -3.357110 & 1.7$\pm$0.1 & 59$\pm$5\\ 
47 & 76.324990 & -3.490043 & 6.0$\pm$2.4 & 123$\pm$11\\ 
48 & 76.330111 & -4.138565 & 0.2$\pm$0.1 & 129$\pm$10\\ 
49 & 76.334563 & -3.512351 & 3.7$\pm$0.7 & 12$\pm$6\\ 
50 & 76.339824 & -3.474758 & 0.3$\pm$0.1 & 112$\pm$10\\ 
51 & 76.352278 & -3.356968 & 3.8$\pm$0.9 & 38$\pm$8\\ 
52 & 76.357064 & -3.476143 & 0.7$\pm$0.1 & 180$\pm$8\\ 
53 & 76.358799 & -3.318961 & 4.9$\pm$0.5 & 118$\pm$3\\ 
54 & 76.383819 & -3.496369 & 2.9$\pm$0.7 & 72$\pm$7\\ 
55 & 76.394011 & -3.252690 & 3.4$\pm$1.0 & 122$\pm$8\\ 
56 & 76.395233 & -3.364700 & 4.0$\pm$0.7 & 98$\pm$5\\ 
57 & 76.395640 & -3.446096 & 0.9$\pm$0.3 & 27$\pm$9\\ 
58 & 76.401360 & -3.359236 & 0.9$\pm$0.1 & 151$\pm$5\\ 
59 & 76.402862 & -3.245707 & 0.9$\pm$0.3 & 110$\pm$8\\ 
60 & 76.404832 & -3.367907 & 0.4$\pm$0.1 & 20$\pm$8\\
61 & 76.406890 & -3.244556 & 3.9$\pm$1.8 & 170$\pm$12\\ 
62 & 76.413576 & -3.256279 & 1.0$\pm$0.2 & 122$\pm$7\\   
\hline
\end{tabular}\\
\end{table}

\begin{table}
\centering
\contcaption{}
\begin{tabular}{lcccr}
\hline
Star &  RA(2016) & Dec(2016) & $P\pm\Delta P$ & $\theta\pm\Delta\theta$ \\
ID  & ($^{\circ}$) & ($^{\circ}$) & (\%) & ($^{\circ}$)\\
(1) & (2) & (3) & (4) & (5)\\
\hline 
63 & 76.422006 & -4.052387 & 0.1$\pm$0.0 & 176$\pm$13\\
64 & 76.425105 & -3.227139 & 0.2$\pm$0.1 & 113$\pm$10\\ 
65 & 76.426136 & -3.299343 & 2.1$\pm$1.0 & 89$\pm$11\\ 
66 & 76.430966 & -4.081323 & 3.6$\pm$0.8 & 112$\pm$6\\ 
67 & 76.432660 & -4.084967 & 3.3$\pm$0.5 & 97$\pm$4\\ 
68 & 76.441285 & -3.466186 & 0.5$\pm$0.1 & 63$\pm$8\\ 
69 & 76.443448 & -3.263562 & 0.6$\pm$0.1 & 105$\pm$5\\ 
70 & 76.446219 & -4.050299 & 0.9$\pm$0.2 & 175$\pm$6\\ 
71 & 76.448053 & -3.220008 & 0.2$\pm$0.1 & 142$\pm$8\\ 
72 & 76.449939 & -3.383702 & 0.8$\pm$0.1 & 101$\pm$5\\ 
73 & 76.451263 & -3.336181 & 4.7$\pm$0.5 & 75$\pm$3\\ 
74 & 76.452056 & -4.096511 & 0.5$\pm$0.2 & 127$\pm$7\\ 
75 & 76.453728 & -3.502021 & 0.7$\pm$0.2 & 114$\pm$7\\ 
76 & 76.456614 & -3.343400 & 2.9$\pm$0.9 & 91$\pm$8\\ 
77 & 76.457071 & -3.382982 & 0.5$\pm$0.2 & 80$\pm$7\\ 
78 & 76.459017 & -3.237327 & 3.1$\pm$0.7 & 75$\pm$7\\ 
79 & 76.459936 & -3.260247 & 0.5$\pm$0.1 & 81$\pm$8\\ 
80 & 76.462164 & -3.327298 & 0.8$\pm$0.2 & 86$\pm$7\\ 
81 & 76.469395 & -3.999568 & 9.1$\pm$1.0 & 175$\pm$3\\ 
82 & 76.472382 & -4.079207 & 0.7$\pm$0.3 & 86$\pm$9\\ 
83 & 76.474180 & -3.840297 & 3.2$\pm$0.3 & 98$\pm$4\\ 
84 & 76.474699 & -3.344415 & 0.3$\pm$0.2 & 129$\pm$9\\ 
85 & 76.487716 & -3.265506 & 2.2$\pm$0.4 & 95$\pm$5\\ 
86 & 76.489946 & -3.241836 & 3.8$\pm$0.8 & 51$\pm$7\\ 
87 & 76.491201 & -3.869157 & 0.2$\pm$0.2 & 149$\pm$12\\ 
88 & 76.492322 & -3.885335 & 0.5$\pm$0.1 & 175$\pm$8\\ 
89 & 76.499055 & -4.013831 & 7.1$\pm$0.6 & 111$\pm$3\\ 
90 & 76.501065 & -4.049929 & 5.1$\pm$1.1 & 0$\pm$6\\ 
91 & 76.501643 & -3.826313 & 4.3$\pm$0.6 & 42$\pm$4\\ 
92 & 76.504388 & -3.467302 & 0.7$\pm$0.3 & 57$\pm$11\\ 
93 & 76.505833 & -3.850232 & 3.1$\pm$0.9 & 18$\pm$8\\ 
94 & 76.508228 & -3.241734 & 3.0$\pm$0.6 & 126$\pm$6\\ 
95 & 76.514145 & -3.508980 & 2.2$\pm$0.9 & 138$\pm$12\\ 
96 & 76.514333 & -3.815288 & 1.2$\pm$0.6 & 154$\pm$13\\ 
97 & 76.516700 & -3.857782 & 6.9$\pm$0.8 & 51$\pm$3\\ 
98 & 76.518382 & -3.909460 & 1.5$\pm$0.6 & 137$\pm$11\\ 
99 & 76.519196 & -3.860419 & 0.3$\pm$0.2 & 162$\pm$12\\ 
100 & 76.522211 & -3.859188 & 0.8$\pm$0.1 & 102$\pm$5\\ 
101 & 76.524547 & -4.059353 & 3.0$\pm$1.0 & 26$\pm$9\\ 
102 & 76.525837 & -3.019217 & 1.9$\pm$1.0 & 12$\pm$13\\ 
103 & 76.528123 & -3.586487 & 2.2$\pm$1.0 & 54$\pm$11\\ 
104 & 76.531684 & -3.473344 & 1.7$\pm$0.2 & 66$\pm$4\\ 
105 & 76.532659 & -3.419636 & 1.9$\pm$0.2 & 59$\pm$4\\ 
106 & 76.533053 & -3.246493 & 1.6$\pm$0.8 & 161$\pm$13\\ 
107 & 76.535788 & -3.889359 & 1.4$\pm$0.4 & 101$\pm$8\\ 
108 & 76.536405 & -3.607700 & 2.6$\pm$0.8 & 10$\pm$9\\ 
109 & 76.538142 & -3.888124 & 0.5$\pm$0.3 & 135$\pm$13\\ 
110 & 76.540154 & -3.270674 & 3.3$\pm$1.0 & 141$\pm$8\\ 
111 & 76.540777 & -3.252771 & 4.1$\pm$0.9 & 96$\pm$7\\ 
112 & 76.541560 & -3.324007 & 4.6$\pm$0.9 & 85$\pm$5\\ 
113 & 76.542465 & -3.268319 & 0.3$\pm$0.2 & 106$\pm$9\\ 
114 & 76.545834 & -3.245438 & 0.7$\pm$0.3 & 95$\pm$10\\ 
115 & 76.547944 & -3.481960 & 2.9$\pm$1.3 & 162$\pm$12\\ 
116 & 76.550182 & -3.809668 & 0.5$\pm$0.1 & 2$\pm$8\\ 
117 & 76.553295 & -3.228827 & 0.1$\pm$0.1 & 145$\pm$13\\ 
118 & 76.558389 & -3.544870 & 1.7$\pm$1.0 & 100$\pm$13\\ 
119 & 76.564795 & -3.892936 & 1.8$\pm$0.8 & 9$\pm$11\\ 
120 & 76.566039 & -3.257079 & 1.5$\pm$0.6 & 175$\pm$16\\ 
121 & 76.566386 & -3.406856 & 1.1$\pm$0.3 & 29$\pm$8\\ 
122 & 76.571028 & -3.822301 & 0.2$\pm$0.1 & 157$\pm$17\\ 
123 & 76.571268 & -3.573746 & 0.7$\pm$0.3 & 41$\pm$10\\
124 & 76.575936 & -3.069034 & 6.6$\pm$0.7 & 96$\pm$3\\ 
125 & 76.576201 & -3.418317 & 0.8$\pm$0.2 & 45$\pm$7\\ 
\hline
\end{tabular}\\
\end{table}

\begin{table}
\centering
\contcaption{}
\begin{tabular}{lcccr} 
\hline
Star &  RA(2016) & Dec(2016) & $P\pm\Delta P$ & $\theta\pm\Delta\theta$ \\
ID  & ($^{\circ}$) & ($^{\circ}$) & (\%) & ($^{\circ}$)\\
(1) & (2) & (3) & (4) & (5)\\
\hline 
126 & 76.576361 & -3.486160 & 3.4$\pm$0.8 & 93$\pm$6\\  
127 & 76.578532 & -3.249101 & 0.6$\pm$0.4 & 92$\pm$12\\ 
128 & 76.579960 & -3.458023 & 0.9$\pm$0.4 & 13$\pm$11\\ 
129 & 76.586041 & -3.257971 & 2.0$\pm$0.4 & 157$\pm$6\\ 
130 & 76.589001 & -3.223807 & 3.9$\pm$0.5 & 66$\pm$4\\ 
131 & 76.589297 & -3.871959 & 0.1$\pm$0.1 & 143$\pm$10\\ 
132 & 76.592087 & -3.302383 & 0.7$\pm$0.3 & 122$\pm$9\\ 
133 & 76.596953 & -3.275525 & 1.9$\pm$0.8 & 104$\pm$10\\ 
134 & 76.604839 & -3.242774 & 0.7$\pm$0.4 & 134$\pm$13\\ 
135 & 76.606203 & -3.266237 & 5.3$\pm$1.7 & 178$\pm$8\\ 
136 & 76.607265 & -3.282274 & 0.7$\pm$0.3 & 7$\pm$11\\ 
137 & 76.608836 & -3.888718 & 6.7$\pm$0.9 & 129$\pm$4\\ 
138 & 76.609029 & -3.227658 & 0.7$\pm$0.1 & 2$\pm$5\\ 
139 & 76.609647 & -3.304808 & 2.8$\pm$1.4 & 154$\pm$13\\ 
140 & 76.612248 & -3.590443 & 5.8$\pm$1.6 & 118$\pm$8\\ 
141 & 76.616531 & -3.501308 & 0.7$\pm$0.3 & 179$\pm$11\\ 
142 & 76.620341 & -3.520941 & 4.1$\pm$1.0 & 33$\pm$6\\ 
143 & 76.623236 & -3.503529 & 4.2$\pm$1.0 & 40$\pm$6\\ 
144 & 76.626958 & -3.923052 & 0.9$\pm$0.3 & 126$\pm$6\\ 
145 & 76.628258 & -3.589061 & 0.2$\pm$0.1 & 162$\pm$10\\ 
146 & 76.628374 & -3.892122 & 2.4$\pm$1.1 & 39$\pm$12\\ 
147 & 76.633333 & -3.610037 & 2.3$\pm$1.4 & 100$\pm$14\\ 
148 & 76.637066 & -3.923354 & 1.4$\pm$0.7 & 98$\pm$12\\ 
149 & 76.637160 & -3.543341 & 3.3$\pm$1.5 & 14$\pm$12\\ 
150 & 76.640089 & -3.466988 & 2.5$\pm$0.7 & 101$\pm$7\\ 
151 & 76.640276 & -3.917883 & 3.5$\pm$1.8 & 103$\pm$13\\ 
152 & 76.643346 & -3.550167 & 0.3$\pm$0.3 & 109$\pm$12\\ 
153 & 76.644272 & -3.910570 & 1.2$\pm$0.4 & 159$\pm$8\\ 
154 & 76.647270 & -3.464998 & 2.7$\pm$1.3 & 155$\pm$22\\ 
155 & 76.657264 & -3.539520 & 2.0$\pm$0.6 & 111$\pm$7\\ 
156 & 76.659403 & -3.425535 & 2.4$\pm$1.0 & 21$\pm$11\\ 
157 & 76.666073 & -3.942928 & 1.4$\pm$0.4 & 35$\pm$8\\ 
158 & 76.679738 & -3.498806 & 0.2$\pm$0.2 & 145$\pm$10\\ 
159 & 76.702961 & -3.476747 & 2.3$\pm$1.1 & 32$\pm$12\\ 
160 & 76.704717 & -3.192237 & 1.0$\pm$0.2 & 140$\pm$6\\ 
161 & 76.720608 & -3.169351 & 0.2$\pm$0.1 & 169$\pm$10\\ 
162 & 76.732255 & -3.238896 & 0.6$\pm$0.2 & 144$\pm$9\\ 
163 & 76.749136 & -3.268334 & 2.0$\pm$0.4 & 25$\pm$7\\ 
164 & 76.749492 & -3.202306 & 4.1$\pm$0.8 & 180$\pm$6\\ 
165 & 76.763398 & -3.231897 & 3.1$\pm$0.7 & 10$\pm$8\\ 
166 & 76.770172 & -3.166013 & 0.2$\pm$0.1 & 113$\pm$10\\ 
167 & 76.778020 & -3.197624 & 2.4$\pm$0.7 & 138$\pm$8\\ 
168 & 76.779425 & -3.514769 & 3.4$\pm$1.4 & 121$\pm$10\\ 
169 & 76.781153 & -3.439858 & 0.6$\pm$0.2 & 140$\pm$7\\ 
170 & 76.782123 & -3.114975 & 0.8$\pm$0.3 & 97$\pm$7\\ 
171 & 76.793102 & -3.430310 & 0.3$\pm$0.1 & 157$\pm$10\\ 
172 & 76.800466 & -3.515781 & 2.9$\pm$1.3 & 153$\pm$11\\ 
173 & 76.803432 & -3.374655 & 2.7$\pm$0.2 & 20$\pm$3\\ 
174 & 76.807782 & -3.204995 & 1.1$\pm$0.4 & 141$\pm$8\\ 
175 & 76.826055 & -3.580222 & 0.6$\pm$0.3 & 46$\pm$11\\ 
176 & 76.826319 & -3.276114 & 2.2$\pm$0.6 & 144$\pm$8\\ 
177 & 76.826512 & -3.442943 & 3.3$\pm$0.4 & 151$\pm$4\\ 
178 & 76.833890 & -3.647206 & 0.9$\pm$0.4 & 101$\pm$10\\ 
179 & 76.834699 & -3.568827 & 0.3$\pm$0.1 & 155$\pm$13\\ 
180 & 76.835544 & -3.321679 & 1.6$\pm$0.4 & 94$\pm$4\\ 
181 & 76.840358 & -3.468366 & 0.7$\pm$0.3 & 130$\pm$8\\ 
182 & 76.844702 & -3.447446 & 1.6$\pm$0.8 & 86$\pm$12\\ 
183 & 76.845020 & -3.503303 & 0.5$\pm$0.2 & 103$\pm$10\\ 
184 & 76.854125 & -3.449802 & 5.9$\pm$1.5 & 106$\pm$7\\ 
185 & 76.859193 & -3.248541 & 3.2$\pm$0.6 & 43$\pm$5\\ 
186 & 76.860937 & -3.477223 & 3.8$\pm$0.6 & 67$\pm$5\\
187 & 76.864791 & -3.302551 & 1.1$\pm$0.5 & 41$\pm$7\\ 
188 & 76.869626 & -3.280521 & 0.9$\pm$0.2 & 37$\pm$13\\ 
\hline
\end{tabular}\\
\end{table}

\begin{table}
\centering
\contcaption{}
\begin{tabular}{lcccr} 
\hline
Star &  RA(2016) & Dec(2016) & $P\pm\Delta P$ & $\theta\pm\Delta\theta$ \\
ID  & ($^{\circ}$) & ($^{\circ}$) & (\%) & ($^{\circ}$)\\
(1) & (2) & (3) & (4) & (5)\\
\hline 
189 & 76.872468 & -3.311369 & 0.3$\pm$0.1 & 96$\pm$8\\ 
190 & 76.880090 & -3.375164 & 0.1$\pm$0.1 & 122$\pm$10\\ 
191 & 76.886432 & -3.649749 & 1.8$\pm$0.3 & 120$\pm$5\\  
192 & 76.890162 & -3.333211 & 0.4$\pm$0.3 & 114$\pm$12\\ 
193 & 76.892951 & -3.548734 & 0.4$\pm$0.3 & 133$\pm$12\\ 
194 & 76.910323 & -3.361275 & 2.9$\pm$0.9 & 10$\pm$9\\ 
195 & 76.920860 & -3.381351 & 1.7$\pm$0.6 & 122$\pm$9\\ 
		\hline
\end{tabular}\\
\end{table}
\begin{table*}
    \begin{center}
    \caption{Properties of the newly found comoving sources identified towards L1616 from \textit{Gaia} EDR3.}
    \label{tab:com_gaia}
    \renewcommand{\arraystretch}{1.3}
    \footnotesize
    \begin{tabular}{lccccccccc} 
    \hline
    \#& RA(2016)  & Dec(2016)  & Source ID& $d(\Delta d)$ & $\mu_{\alpha\star}$ ($\Delta\mu_{\alpha\star}$) & $\mu_{\delta}$ ($\Delta\mu_{\delta}$) & $G(\Delta G)$ & $G_\mathrm{BP}(\Delta G_\mathrm{BP})$ & $G_\mathrm{RP}(\Delta G_\mathrm{RP})$\\
     & ($^{\circ}$) & ($^{\circ}$) & \textit{Gaia} EDR3 & (pc)& (mas yr$^{-1}$) & (mas yr$^{-1}$) & (mag)& (mag) & (mag)\\
     (1)&(2)&(3)&(4)&(5)&(6)&(7)&(8)&(9)&(10)\\
    \hline
    c1 & 76.872468 & -3.311369 & 3213013808064290304 & 380$^{4}_{-4}$ & 1.174$\pm$0.030 & -1.068$\pm$0.021 & 8.862$\pm$0.003 & 8.834$\pm$0.003 & 8.868$\pm$0.004\\
    c2 & 76.991670 & -3.589467 & 3212940067769419776 & 380$^{24}_{-18}$ & 1.095$\pm$0.149 & -1.501$\pm$0.115 & 18.243$\pm$0.003&20.550$\pm$0.074 & 16.821$\pm$0.009\\    
    \hline
    \end{tabular}\\
    \end{center}
\end{table*}

\section{Figures}

\begin{figure*}
	\includegraphics[width=5.9cm, height=5.3cm]{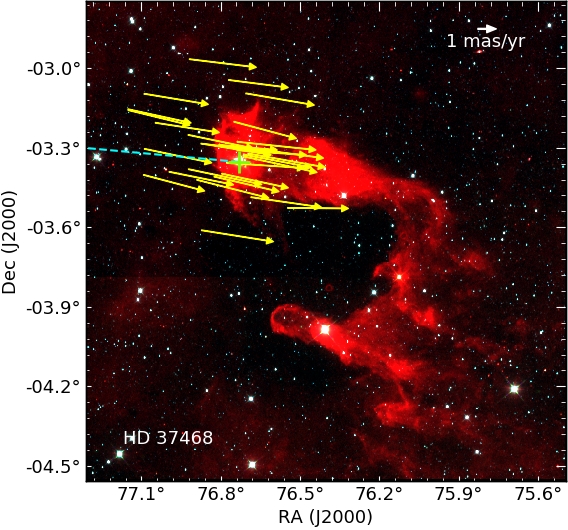}
	\includegraphics[width=5.9cm, height=5.3cm]{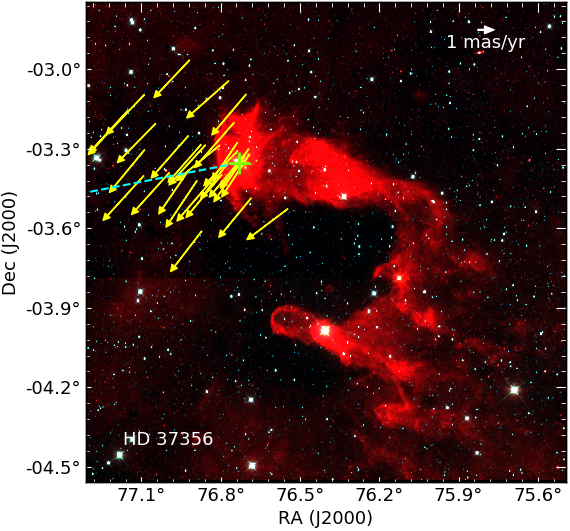}
	\includegraphics[width=5.9cm, height=5.3cm]{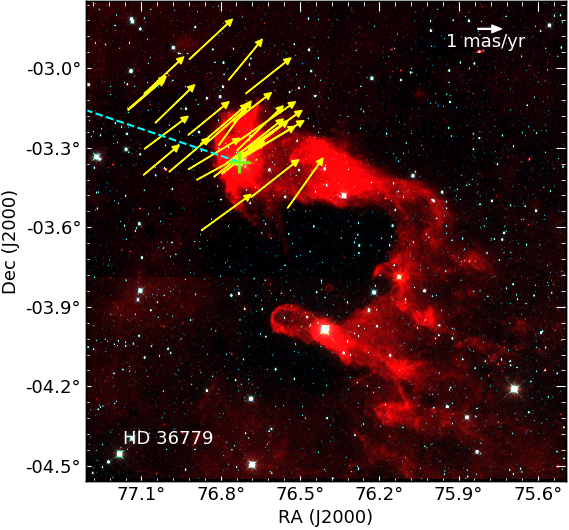}
	\includegraphics[width=5.9cm, height=5.3cm]{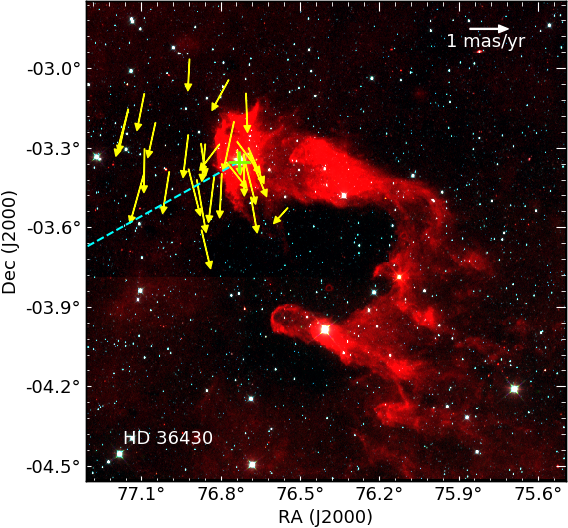}
	\includegraphics[width=5.9cm, height=5.3cm]{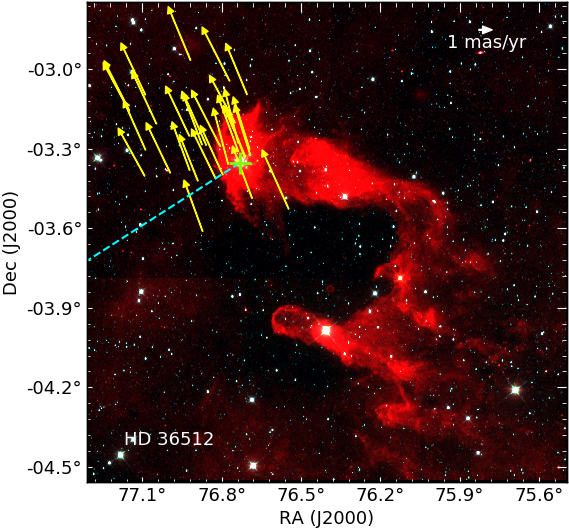}
	\includegraphics[width=5.9cm, height=5.3cm]{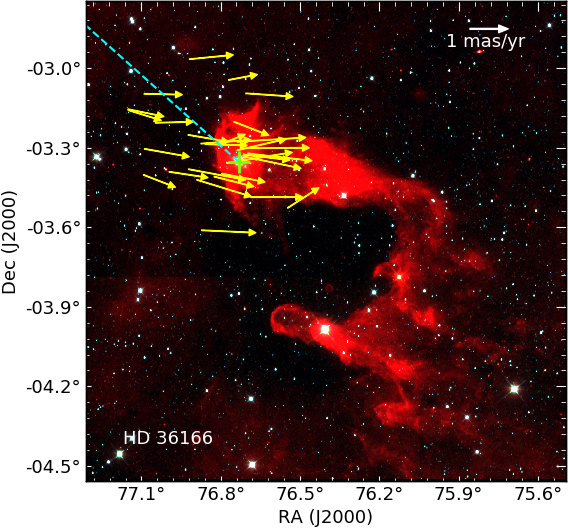}
	\includegraphics[width=5.9cm, height=5.3cm]{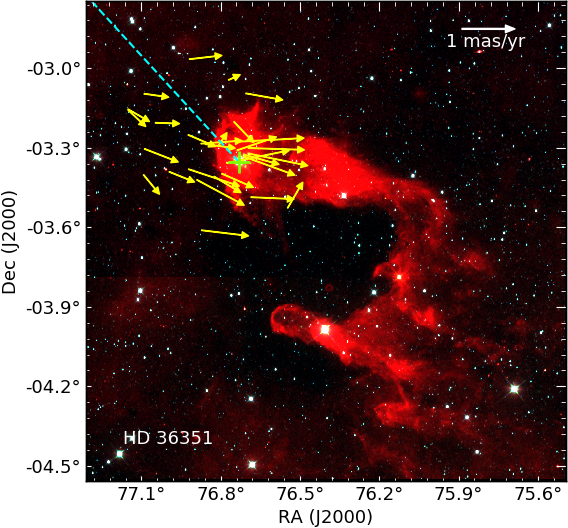}
	\includegraphics[width=5.9cm, height=5.3cm]{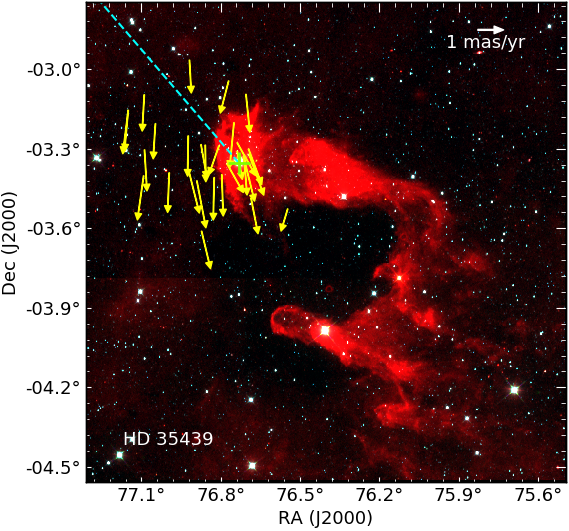}	
	\includegraphics[width=5.9cm, height=5.3cm]{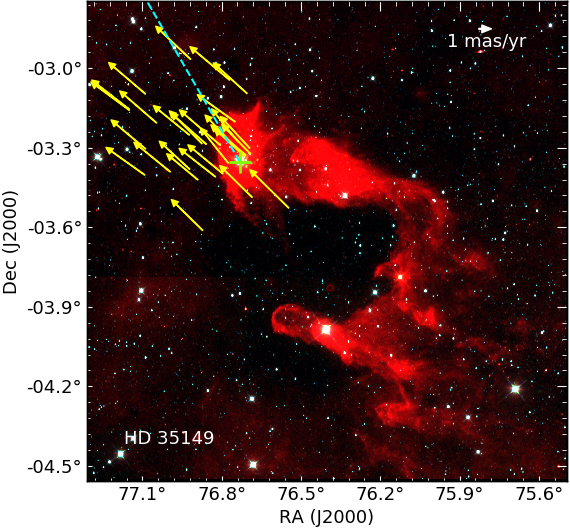}	
	\caption{The relative proper motion vectors of the known YSOs with respect to the other massive sources (listed in table \ref{tab:ob_stars_table}) are shown using yellow vectors. The names of the massive stars are labelled in the bottom left corner of the images. The directions of the massive stars from the \textit{IRAS} source (green `+' symbol) embedded in L1616 are indicated by dashed cyan lines.The proper motion values of HD 37468 ($\sigma$ Ori) are acquired from \textit{Hipparcos}. The same for other sources are retrieved from \textit{Gaia} EDR3 catalog.}
	\label{fig:pm_sigma}
\end{figure*}

\bsp	
\label{lastpage}
\end{document}